%% file: GB_JPG_2019_v8_final.tex
\documentclass[12pt]{iopart}

\usepackage{graphicx}
\usepackage{hyperref}
\hypersetup{
  colorlinks,%
  citecolor=black,%
  filecolor=black,%
  linkcolor=black,%
  urlcolor=blue
  }
  
\usepackage[noadjust]{cite}
\usepackage{float}
\usepackage{subcaption}
\usepackage{multirow}
\usepackage{iopams}

\usepackage{bm}

\newcommand{\GeV}{\ \mathrm{GeV}}

\newcommand{\dd}{ {\mathrm d} } 

\begin{document}

\title[Tsallis-thermometer: a QGP indicator for large and small collisional systems]{Tsallis-thermometer: a QGP indicator for large and small collisional systems}

\author{G\'abor B\'ir\'o$^{1,2}$, Gergely G\'abor Barnaf\"oldi$^1$ and Tam\'as S\'andor Bir\'o$^1$}
  
\address{$^1$Wigner Research Center for Physics, 29--33 Konkoly--Thege Mikl\'os Str., H-1121 Budapest, Hungary.}
\address{$^2$Institute of Physics, E\"otv\"os Lor\'and University, 1/A P\'azm\'any P\'eter S\'et\'any, H-1117  Budapest, Hungary.}

\ead{biro.gabor@wigner.hu; barnafoldi.gergely@wigner.hu; biro.tamas@wigner.hu}

\begin{indented}
  \item[]\today
  \end{indented}

\begin{abstract}
  The transverse momentum distribution of identified hadrons from recent years are analyzed within the thermodynamically consistent formulation of non-extensive statistics. A wide range of center-of-mass energies and average event multiplicities are studied for various hadron species.  
  We demonstrate that the average event multiplicity is a key variable in the study of high-energy collisions and a smooth transition in the description from small to large systems is possible. For this purpose the non-extensive statistical approach is more than appropriate. 
  The validity of the non-extensive description is explored in multiple ways: such as the calculation of integrated yields per unit rapidity, the analysis of radial flow and the consistency check of the thermodynamical variables. The 'Tsallis-thermometer' is introduced as an indicator of quark-gluon plasma in small collisional systems.
\end{abstract}


\vspace{2pc}

\submitto{\JPG}


%
%

\section{Introduction}
\label{sec:introduction}

The quark-gluon plasma (QGP) is believed to form in high-energy heavy-ion collisions, once the energy density is high enough. The properties of the QGP can be studied through transverse momentum distributions of identified hadrons. These distributions are measured by complex detector systems such as the ALICE, CMS, ATLAS and LHCb at the LHC (CERN, Switzerland) or PHENIX, STAR, BRAHMS and PHOBOS at the RHIC (BNL, USA). Recently, phenomena, interpreted as being the sign of QGP formation, were measured in small collisional systems, such as pPb and high-multiplicity pp collisions~\cite{ALBERTA:collecticity, GrosseOetringhaus:2020zhx, ALICE:2017jyt, Abelev:2012ola, Aad:2012gla, Khachatryan:2010gv, PHENIX:2018lia, Loizides:2016tew, Mishra:2019png, Varga:2018isd, Mishra:2018pio, Kanakubo:2019ogh, Biro:1994mp}. With the discovery of the strong correlations in these systems the dominance of the hydrodynamic description became uncertain~\cite{Loizides:2016tew}. Since the measured collective behavior of particles is assumed to be the consequence of the expanding hot and dense matter, characteristic only to heavy-ion collisions so far, the appearance of similar phenomena raises many new questions regarding the QGP. Meanwhile also an increase of strangeness production was measured in small systems recently: a signature often connected to deconfinement~\cite{ALICE:2017jyt}. At the same time, this enhancement was observed to be surprisingly smooth as a function of charged particle multiplicity, covering a broad range of pp, pA and AA collisions. These measurements induced a paradigm shift in high-energy heavy-ion physics, indicating the growing importance of small systems and the necessity of their continuous connection to the large, heavy-ion systems.

The spectra of identified hadrons incorporate information about the ongoing parton-level processes during the hadronization. Hadron formation is still a faintly understood phenomenon, especially because of its non-perturbative nature. The high-$p_T$ part of the spectrum corresponds to the pQCD calculable jet-like origin, while the soft/bulk part is traditionally considered in terms of thermodynamics and hydrodynamics~\cite{Ollitrault:1992bk, BraunMunzinger:2003zd, Cleymans:1992zc}. On the other hand, the development of a unified model that explains both regions simultaneously is a widely studied topic, with several different approaches~\cite{Molnar:2019yam, Takacs:2019ikb, Shi:2018vys, Damodaran:2017ior}. Especially in small systems at high multiplicity the hardening of the spectra (and possible a hint for jet modification) was observed. Here the application of the non-extensive statistical framework seems to be straightforward~\cite{Mishra:2018pio, Mishra:2019png, Varga:2018isd, Varga:2018enp, Varga:2019rhi}.

It is a well known fact that Tsallis\,--\,Pareto distributions connect the power-law tailed and exponential-like (i.e. Boltzmann\,--\,Gibbs) distributions in a smooth way~\cite{artic:tsorig, book:ts2, BG:entr17, BG:maxent16}. Although the development of a convincing physical explanation is still in progress, the application of these distributions to high-energy nuclear collisions is very successful~\cite{artic:tsbphysica, artic:flowing0, artic:gbiro, artic:cleymansphyslett, artic:cleymansjphys, artic:cleymansjphys2, artic:wilk1, artic:wilk2, artic:wilk3, artic:wilk4, Wilk:2015pva, artic:actaphys, artic:chaossolitions, artic:lilin1, artic:lilin2, CL:jphys17, Rybczynski:2014ura}. In recent studies we have shown that for hadron yields with negative binomial multiplicity distributions it is not only possible but definitely necessary to assume a cut power-law distribution for the one-particle energy~\cite{artic:tsbphysica, artic:tsbphysica2, artic:tsbeurphys, artic:tsbeurphys2, artic:tsbentr, TSB:EPJ12, book:tsb}. In high-energy particle collision experiments with high collision rate, where both the number of the produced particles and the energies of these particles have large event-by-event fluctuations, it is practically very difficult to measure every aspects of an event~\cite{Levai:2003at, Agocs:2011sda, Acconcia:2013ptg}. In real data analysis practically the statistical average of such observables is measured, taking into account as many events as possible. This leads to a natural emergence of the Tsallis\,--\,Pareto distribution for the transverse momentum. 

In this paper we provide a comprehensive picture behind the non-extensive parameters extracted from recent datasets~\cite{Aamodt:2011zza, Abelev:2012jp, Song:2017gnh, Adams:2006nd, Abelev:2006pp, Adams:2003qm, Adam:2015gka, Adam:2015qaa, Chatrchyan:2013eya, Adcox:2001mf,Adam:2016dau,Abelev:2013haa,Adam:2016bpr,Adamova:2017elh,Adam:2015vsf,ALICE:2017jyt, Abelev:2013vea, Adare:2011vy, Aamodt:2011zj,Abelev:2014laa, Arsene:2005mr, Adam:2017zbf, Abelev:2013xaa, ABELEV:2013zaa, Acharya:2019kyh, Acharya:2019yoi, Agakishiev:2011ar, Abelev:2008zk, Back:2002uc, Alver:2010ck, Abelev:2013qoq, Acharya:2018orn, Chatrchyan:2012qb}. Thanks to the rapid technological development of the detector and accelerator instrumentation, identified hadron spectra with better statistics than ever are available, even with multiplicity selection. By analyzing these spectra including the power-law tail it is possible to reveal connections between hadron production and observations regarding collective effects from large to small systems. Moreover we are able to unveil non-trivial correlations between the Tsallis parameters and the center-of-mass energy, $\sqrt{s}$, the hadron mass, and the event multiplicity~\cite{ALICE:2017jyt, Cleymans:2011in, Santos:2014ota, Deppman:2012qt, Megias:2015fra, artic:gribov80, Cleymans:2013tna, ALBERTA:collecticity, Rybczynski:2014ura, Grigoryan:2017gcg, Parvan:2016rln, Ortiz:2016kpz, Yin:2017pzp, Tripathy:2016hlg, Biro:2016dgd}. 

Our paper is organized as follows. First we give a brief theoretical summary on the non-extensive statistical approach in \sref{sec:nonext}. Then in \sref{sec:expdata} we list the datasets we used in this study. In \sref{sec:evolution} we discuss the scaling behavior of the resulted Tsallis\,--\,Pareto parameters versus center-of-mass energy, $\sqrt{s}$, hadron mass, and multiplicity. In \sref{sec:flow} we investigate the radial flow and freeze-out temperature in the context of the non-extensive statistics and compare our results with other works. In \sref{sec:thermo} the corresponding thermodynamical quantities are shown, such as the entropy density, $s$, energy density, $\varepsilon$, and pressure, $P$. The discussion and interpretation of our results and a summary follow in \sref{sec:discussion} and \sref{sec:summary}. Finally, the detailed lists of the experimental datasets with references and fit parameters are presented in the Appendices. For convenience, in this paper we apply $c=\hbar=k_B=1$ units.

\section{Introduction to the non-extensive statistical framework}
\label{sec:nonext}

The most studied observables in high-energy particle collisions are the transverse momentum distributions of the produced, identified hadrons. During the several decades of their study, the thermal description has gathered quite a few achievements: the assumption that the final state particles are stemming from a system in thermal and chemical equilibrium in the hadronic phase long seemed to be adequate~\cite{BraunMunzinger:2003zd, Cleymans:1992zc}. According to the Boltzmann\,--\,Gibbs statistics, in this model the hadron spectra is described by exponential distributions in the form
\begin{equation}
  f(E)\sim \exp\left(-\frac{E-\mu}{T}\right),
  \label{eq:BG}
\end{equation}
where $\mu$ is chemical potential and $T$ is an associated temperature. However, it turned out that exponential fits are failing at the moderate and high-$p_T$ regime ($p_T \gtrsim 3$ GeV) of the spectra. They rather follow a power-law tailed distribution. This was attributed to the assumption that in this regime the hadron production is governed by physically different, non-thermal and non-equilibrium perturbative QCD processes. It is worthwhile to note that the difference in the yield of the two, separate regimes can reach even several orders of decimal magnitude.

In order to connect these two parts in a meaningful way, an improvement in the statistical picture is unavoidable. This can comfortably be done in the context of non-extensive statistics. This discipline was originally formulated more than three decades ago as an attempt to generalize the Boltzmann\,--\,Gibbs theory~\cite{artic:tsorig, book:ts2}. Though the theory has many far-reaching consequences, the essence can be comprised in the use of a particular generalization of the exponential and logarithm functions:
\begin{eqnarray}
  \exp_q (x) & = \left[1+(1-q)x\right]^{1/(1-q)},
  \label{eq:qexp}
\end{eqnarray}
called '\textit{q-exponential}', and its inverse function, the '\textit{q-logarithm}':
\begin{eqnarray}
  \ln_q (x) & = \frac{x^{1-q}-1}{1-q}.
  \label{eq:qlog}
\end{eqnarray}
The parameter $q$ is usually called the \textit{measure} of non-extensivity or Tsallis-$q$ parameter, and plays an unique role in the theory. In the $q\rightarrow 1$ limit of this generalization, the usual exponential and logarithm functions are restored:
\begin{eqnarray}
  \lim_{q\rightarrow 1}\exp_q (x) & = \exp (x), \\
  \lim_{q\rightarrow 1}\ln_q (x) & = \ln (x).
\end{eqnarray}
The distributions with the form \eref{eq:qexp} are usually called \textit{Tsallis} or \textit{Tsallis\,--\,Pareto} distributions, named after Constantino Tsallis and Vilfredo Pareto~\cite{artic:tsorig, artic:pareto}. They have a distinctive property that makes them very suitable for studying hadron transverse momentum distributions: they describe the low- and high-$p_T$ limits simultaneously. In the last two decades numerous studies investigated this fact both theoretically and experimentally~\cite{Biro:2005uv, Cleymans:2012ya, artic:tsbentr, artic:tsbphysica, TSB:EPJ12, Cleymans:2008mt, artic:flowing0, artic:gbiro, Shen:2017pyo, Yin:2017pzp, Wilk:2015pva, Abbas:2013bpa, artic:cleymansphyslett, artic:cleymansjphys, artic:cleymansjphys2, artic:wilk1, artic:wilk2, artic:lilin1, artic:lilin2, artic:wilk3, artic:actaphys, artic:chaossolitions, artic:wilk4, Wilk:2008ue, CL:jphys17, artic:tsbphysica2, artic:tsbeurphys2, book:tsb, BG:entr17, BG:maxent16, Biro:2017eip, Biro:2019ngd, Khuntia:2017ite, Bhattacharyya:2017hdc, Urmossy:2009jf, Cleymans:2011in, Santos:2014ota, Deppman:2012qt, Megias:2015fra, artic:gribov80, Cleymans:2013tna, Parvan:2016rln, Grigoryan:2017gcg, Tripathy:2016hlg, Rybczynski:2014ura, Biro:2016dgd, Rybczynski:2014cha, Aamodt:2010pb, Zaccolo:2015udc}.

It is not just an \textit{a posteriori} assumption to use Tsallis\,--\,Pareto distributions for studying transverse momentum distributions of hadrons. The main motivation for the generalized non-extensive statistics was to give a comprehensive and consistent description for probabilistic systems, whose elements have strong, long-range correlations whence the non-additivity of the entropy arises. Since the main object of high-energy heavy-ion physics, the quark-gluon plasma is a system where the presence of strong correlation effects are clearly expected, and -- somewhat surprisingly -- similar effects have been observed also in smaller systems, it is well established to apply a non-extensive formulation for studying the hadronization.

During the application of this to high-energy physics we demand thermodynamical consistency by prescribing the first law of thermodynamics with a first order Euler equation:
\begin{equation}
    P=Ts+\mu n - \varepsilon.
    \label{eq:consistency}
\end{equation}
The the following relations must be satisfied between the pressure $P$, entropy density $s$, particle density $n$, energy density $\varepsilon$ and the volume of the system $V$:
\begin{eqnarray}
    P&=-\left.\frac{\partial\Omega}{\partial V}\right|_{T,\mu}=-\frac{\Omega}{V}, \qquad n&=\left.\frac{\partial P}{\partial \mu}\right|_{T,V}, \qquad s=\left.\frac{\partial P}{\partial T}\right|_{V,\mu}, \\
    \varepsilon& = T \frac{\partial P}{\partial T} + \mu\frac{\partial P}{\partial \mu} -P, \qquad N&=\frac{\partial\Omega}{\partial V}, \qquad S=\frac{\partial\Omega}{\partial T}.
\end{eqnarray}
The grand canonical potential, $\Omega$ is given as the logarithm of the partition function:
\begin{equation}
    \Omega(T,V,\mu)= - T \ln{\mathcal{Z}}
\end{equation}
and assuming spatial homogeneity in the volume $V$:
\begin{equation}
    \ln\mathcal{Z}=V\int\frac{\dd^3p}{(2\pi)^3} f\left(\frac{E-\mu}{T}\right).
\end{equation}
As a next step, we assume that the function $f$ is of Tsallis\,--\,Pareto-type \eref{eq:qexp} instead of the conventional Boltzmann\,--\,Gibbs distribution \eref{eq:BG}:
\begin{equation}
  f(E, q, T_q, \mu) = \left[1+\frac{q-1}{T_q}\left(E -\mu \right)\right]^{-\frac{1}{q-1}},
  \label{eq:ts2}
\end{equation}
where $T_q$ is a temperature following the non-extensive statistics and therefore not necessarily the same as the temperature $T$ in \eref{eq:BG}. Note that in the further parts of this paper we omit the index $q$ of $T_q$ and simply use the notation $T$, whose physical meaning is well known in the $q\rightarrow 1$ Boltzmann\,--\,Gibbs limit. On the other hand, in this analysis we investigate hadron yields measured only at mid-rapidity from RHIC and LHC energies, where $\mu$ is small. For the relativistic case, the approximation $\mu\approx m$ is assumed, where $m$ is the rest mass of the given hadron species.

Therefore, based on \label{eq:ts2} the following integrals will be used to calculate the thermodynamical variables (neglecting the quantum statistical corrections):
\begin{eqnarray}
  P & = g \int \frac{\dd^3p}{(2\pi)^3}  T f, \label{eq:P} \\
  N & = nV = gV \int \frac{\dd^3p}{(2\pi)^3} f^q, \label{eq:N} \\
  s & = g \int \frac{\dd^3p}{(2\pi)^3} \left[\frac{E-\mu}{T} f^q  +f \right], \label{eq:s} \\
  \varepsilon & = g \int \frac{\dd^3p}{(2\pi)^3} E f, \label{eq:e}
\end{eqnarray}
where $g$ is the degeneracy factor for the given type of degrees of freedom. From \eref{eq:N} the invariant yield follows as being
\begin{equation}
  E\frac{\dd^3N}{\dd p^3}  = \frac{\dd^2N}{p_T \dd p_T \dd y \dd \phi} =gV\cdot E\cdot f^q.
\end{equation}
Finally, taking the $y\approx 0$ mid-rapidity limit, $E=m_T=\sqrt{p_T^2+m^2}$, and using $\mu\approx m$ we arrive at the widely used fitting form for the invariant transverse momentum distribution:
\begin{equation}
  \left.\frac{\dd^2N}{2\pi p_T \dd p_T \dd y}\right|_{y\approx0}= A m_T \left[1+\frac{q-1}{T}(m_T-m) \right]^{-\frac{q}{q-1}}.
  \label{eq:TS}
\end{equation}
Here $A=C_y gV\frac{1}{(2\pi)^3}$ is a normalizing factor and $C_y$ is a shorthand notation for a correction factor for those experimental data, where in the yield the $\eta$ pseudorapidity is given instead of the $y$ rapidity:
\begin{equation}
    \frac{\dd^2N}{\dd p_T \dd \eta} = \frac{p_T}{m_T} \frac{\dd^2N}{\dd p_T \dd y},
\end{equation}
therefore $C_y=m_T/p_T$ in case of pseudorapidity differential yields, and $C_y=1$ for rapidity differential yields. It is important to note that for similar studies in recent years also slightly different variations of \eref{eq:TS} have been used, based on different physical assumptions. The difference of these variations have been studied in details e.g. in \cite{Shen:2019oyu, Shen:2019kil}. It was concluded that all versions of the Tsallis\,--\,Pareto fitting functions give a consistent result within the accuracy of the available experimental data.
Parameters of two different functions of particle energy, $E$, can only be equal at a given energy, but not for all.
Thus there is no selection rule arising from the latest data for the function type. In the present analysis, we use \eref{eq:TS} to fit the experimental data, because it results in a good $\chi^2/ndf$ and it is thermodynamically consistent.

\subsection{The physical origin of the parameters $T$ and $q$}
\label{subsec:Tqpars}
The main goal of this study is to investigate the characteristics of identified hadron spectra using the non-extensive statistics described above, and to reveal the connections behind the Tsallis parameters and the kinematics defined by the collision experiments. Our starting point is the experimental observation that the charged multiplicity of high-energy collisions is dependent on the center-of-mass energy~\cite{Aamodt:2010pb, Abbas:2013bpa}:
\begin{equation}
    \frac{\left<\dd N_{ch}/\dd\eta\right>}{\left<N_{part}\right>/2} \propto
    \left\{\begin{array}{ll}
      s_{NN}^{0.15} & \textrm{for AA,} \\
      s_{NN}^{0.11} & \textrm{for pp.} 
      \end{array}\right.
    \label{eq:Nmult}
\end{equation}
Furthermore, in earlier studies it was demonstrated that in minimum bias proton-proton collisions both parameters $T$ and $q$ show a mild increase with the center-of-mass energy that can be described with a simple pQCD-inspired scale evolution \textit{ansatz}~\cite{artic:tsbeurphys2, BG:entr17}:
\begin{eqnarray}
    T&=T_1 + T_2 \ln{\frac{\sqrt{s}}{m}}, \label{eq:TCM} \\
    q&=q_1 + q_2 \ln{\frac{\sqrt{s}}{m}}. \label{eq:qCM}
\end{eqnarray}
These observations suggests that a relation between the event multiplicity and the Tsallis parameters should be established.

On the other hand, there are studies that revealed some correlation between the $q$ and $T$ parameters~\cite{artic:tsbentr, Biro:2016dgd, Wilk:2012zn, Wilk:1999dr}. This can be measured by the so called 'Tsallis-thermometer' on  the $T$-$q$ diagram. It was shown that the Tsallis\,--\,Pareto distributed transverse momentum leads to a total charged hadron multiplicity, that follows negative binomial distribution (NBD). This is also supported by experimental data, resulting in an NBD parameter $k\sim\mathcal{O}(10)$~\cite{Biro:2016one, Adare:2008ns, Aamodt:2010pp, Khachatryan:2010nk, Aad:2010ac}. This observation results in an interpretation in which the fluctuations of the number of the produced particles, $n$ is taken into account in a one dimensional relativistic gas~\cite{artic:tsbentr}:
\begin{eqnarray}
  T&=\frac{E}{\left<n\right>},  \label{eq:fluctT}\\
  q&= 1-\frac{1}{\left<n\right>}+\frac{\Delta n^2}{\left<n\right>^2}. \label{eq:fluctq}
\end{eqnarray}
Many recent studies showed that the event-by-event multiplicity measurements (and hence their fluctuations) have important and interesting role in high-energy physics, manifested e.g. in the continuous enhancement of (multi)strange hadrons as a function of multiplicity, or the emergence of long-range correlations in high multiplicity pp and pPb events~\cite{ALICE:2017jyt, ALBERTA:collecticity, GrosseOetringhaus:2020zhx, Mishra:2019png, Varga:2018isd, Mishra:2018pio,  Khuntia:2017ite, Zaccolo:2015udc, Abelev:2012ola, Aad:2012gla, Khachatryan:2010gv}. Consequently, in our study we put an emphasis on values of the parameters $T$ and $q$ at different multiplicity classes as well. 

In this framework $T$ and $q$ are intertwined and the connection between them is based on the thermodynamical picture. Equations \eref{eq:fluctT} and \eref{eq:fluctq} give us an opportunity to test the correlations between $T$ and $q$, assuming that the relative size of multiplicity fluctuations are constant:
\begin{eqnarray}
  \frac{\Delta n^2}{\left<n\right>^2}&:=\delta^2,  \label{eq:fluctpar}
\end{eqnarray}
leads to a linear connection in this approximation
  \begin{eqnarray}
    T & = E\left(\delta^2 - (q-1)\right). \label{eq:Tvsqfit}
\end{eqnarray}
Therefore, by measuring the values of $T$ and $q$ we can get closer to identify the thermodynamical regime in the data, which is the main goal of this work.

\section{Experimental datasets}
\label{sec:expdata}

In the present study we analyzed the publicly available mid-rapidity datasets of high-energy proton-proton (pp), proton-nucleus (pA) and nucleus-nucleus (AA) collisions measured by the ALICE, CMS, BRAHMS, STAR and PHENIX collaborations~\cite{Aamodt:2011zza, Abelev:2012jp, Song:2017gnh, Adams:2006nd, Abelev:2006pp, Adams:2003qm, Adam:2015gka, Adam:2015qaa, Chatrchyan:2013eya, Adcox:2001mf,Adam:2016dau,Abelev:2013haa,Adam:2016bpr,Adamova:2017elh,Adam:2015vsf,ALICE:2017jyt, Abelev:2013vea, Adare:2011vy, Aamodt:2011zj,Abelev:2014laa, Arsene:2005mr, Adam:2017zbf, Abelev:2013xaa, ABELEV:2013zaa, Acharya:2019kyh, Acharya:2019yoi, Agakishiev:2011ar, Abelev:2008zk, Back:2002uc, Alver:2010ck, Abelev:2013qoq, Acharya:2018orn, Chatrchyan:2012qb}. The properties of each dataset including the publication reference is summarized for minimum bias pp yields in \tref{tab:exp_sets_1} and for various multiplicity classes measured in pp, pA, and AA collisions in \tref{tab:exp_sets_2} in \ref{apx:expdata}.

It is important to realize that in this study small-to-large system experimental data with a very broad range of selection criteria were investigated simultaneously. As one can ascertain from tables \ref{tab:exp_sets_1} and \ref{tab:exp_sets_2}, many different hadron species and kinematical setups were included in this study. The different experimental methods (e.g. various detector techniques) can also cause slightly different parameter values when comparing all dataset together.

\Tref{tab:exp_sets_1} lists the yields of charged pions, kaons and (anti)protons stemming from pp collisions at various energies. As we have shown in \cite{BG:maxent16, BG:entr17, Biro:2017eip}, the Tsallis\,--\,Pareto parameters of the multiplicity-integrated, minimum bias yields of the identified hadrons are proportional to the logarithm of the center-of-mass energy and a hierarchical ordering with respect to the hadron mass can also be observed.

On the other hand, the minimum bias pp yields are mostly corresponding to low-multiplicity events. In order to resolve the effect of the size of the medium that is created during the collisions, it is necessary to analyze different multiplicity classes. In \tref{tab:exp_sets_2} the investigated yields with various multiplicity classes are enumerated. In order to make a meaningful comparison of the results, the centrality dependent data were converted into multiplicity classes where needed~\cite{Abelev:2013qoq,Acharya:2018orn}. The conversions were performed according to Glauber model calculations that are detailed in the cited experimental papers. Furthermore, these datasets share the common feature that they were measured at mid-rapidity -- although the exact $y$ and $\eta$ ranges differ for each experiment, this will not affect our investigation of the multiplicity dependence significantly. A similar remark can be made for the transverse momentum range of the measured hadron species: albeit the measured highest $p_T$ values vary among  almost all of the investigated spectra, yet a valid global tendency can be observed in the results.

\section{Evolution of the non-extensive parameters}
\label{sec:evolution}

Our ultimate goal is to provide the thermodynamics in this framework together with the entropy production, for that we reveal the connections between the center-of-mass energy, the system size, event multiplicity and hadron mass. Utilizing our earlier findings in \cite{BG:entr17, BG:maxent16}, we extract the Tsallis\,--\,Pareto parameters defined by \eref{eq:TS} and test the relations presented in the previous section.

\subsection{Fitting procedure and performance}
\label{subsec:fitproc}
In order to extract the Tsallis parameters from a broad variety of datasets consistently, we applied the \verb"LMFIT": Non-Linear Least-Squares Minimization and Curve-Fitting for Python package~\cite{matt_newville_2019_3588521}. The fitting procedure is implemented according to the method we have described previously in \cite{BG:entr17, BG:maxent16}: after an initial approximation of the parameters the best parameter values were obtained by separately fitting the high- and low-$p_T$ regions (where the soft/hard separation scale is $p_{T_{sep}}\approx 3 \GeV$), followed by a last fit for the whole $p_T$ region. Here the initial values of the parameters are the results from the previous step. This method is found to be stable and provides the best $\chi^2/ndf$ values, ranging from 0.02 to 14.7. For the precise values and errors of the fitted parameters please consult the tables in \ref{apx:parameters}.

\begin{figure}[H]
    \centering
    \begin{subfigure}[b]{0.49\textwidth}
    \includegraphics[width=\textwidth]{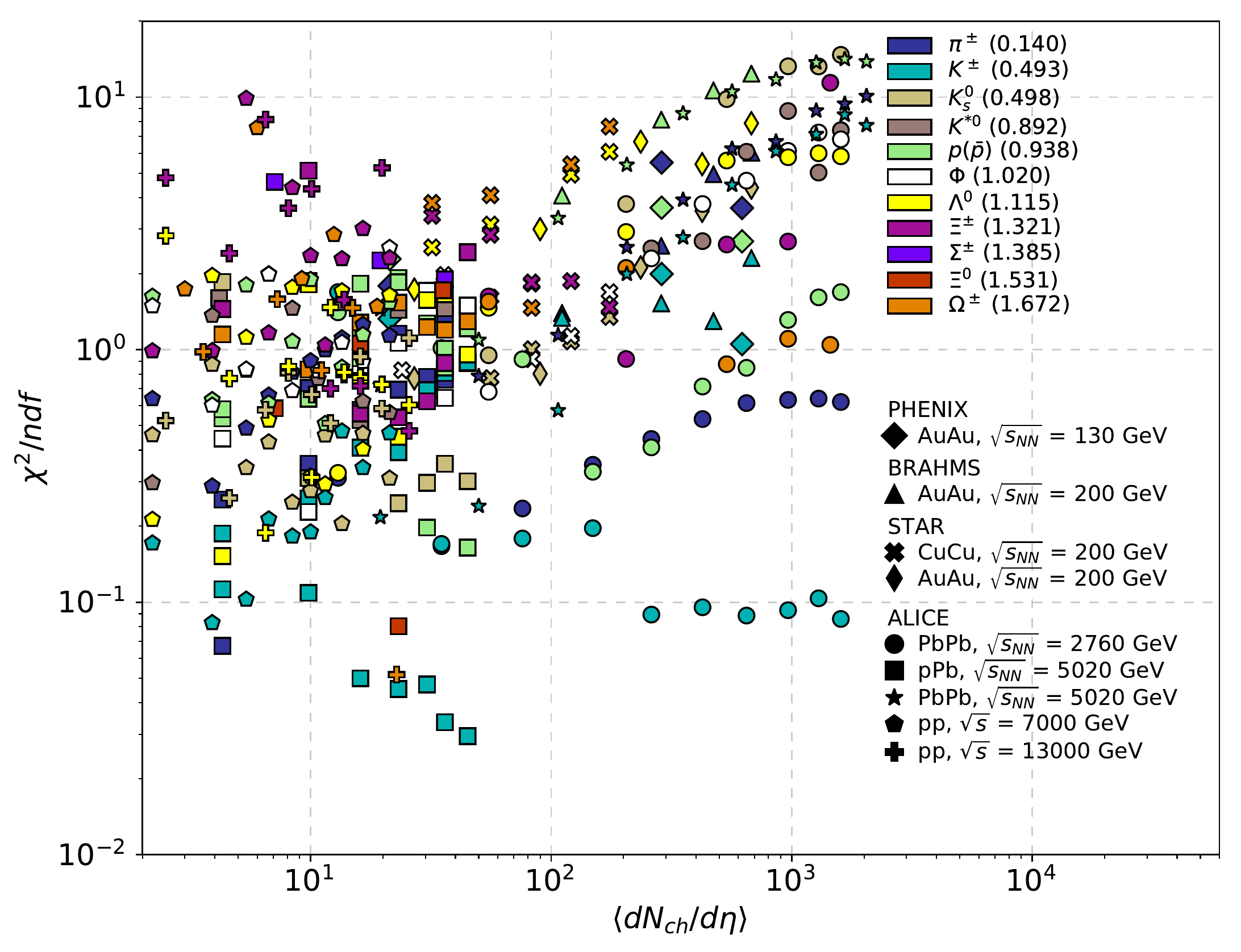}
      \caption{}
    \label{fig:chi2all}
    \end{subfigure}
    \begin{subfigure}[b]{0.49\textwidth}
      \includegraphics[width=\textwidth]{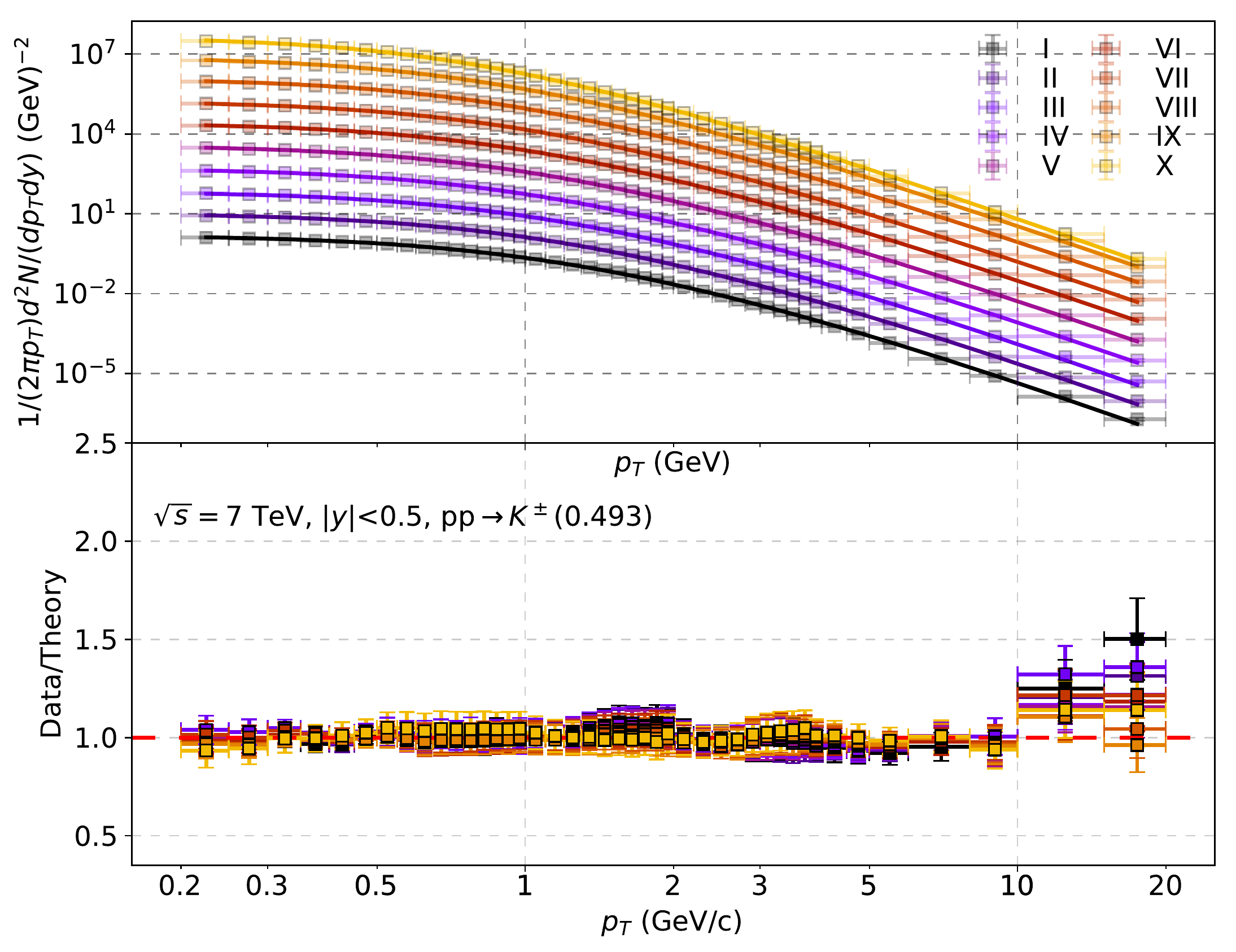}
      \caption{}
      \label{fig:example}
    \end{subfigure}
    \caption{\textit{Left panel:} The $\chi^2/ndf$ values for identified hadrons measured in pp, pA and AA collisions at various collision energies, as function of the event multiplicity. \textit{Right panel:} The yield of charged kaons measured at $\sqrt{s}=7$ TeV pp collisions at different multiplicity classes and the corresponding fitted Tsallis\,--\,Pareto distributions. The roman number '\textrm{I}' denotes the highest multiplicity class, while '\textrm{X}' is the lowest multiplicity class (see text and \tref{tab:exp_sets_2} in \sref{apx:expdata}). For better visibility the different multiplicity classes are shifted.}
    \label{fig:mult_chi2}
\end{figure}

In order to get a global picture on the performance of the fits, the $\chi^2/ndf$ values of all fits as a function of the average event multiplicity is plotted on \fref{fig:mult_chi2}. The different colors represent different hadron species, while the different markers stand for the given CM energy, colliding system and the corresponding experiment (see legend of the figure). As \fref{fig:mult_chi2} presents, the fits are acceptable in all the investigated systems -- however, it gets slightly worse towards the higher multiplicity classes (i.e. at the most central nucleus-nucleus collisions), reaching a value of 14.7 for the most central neutral kaons, stemming from PbPb at $\sqrt{s_{NN}}=2.76$ TeV. An example is shown on \fref{fig:example}, where the yield of charged kaons measured at $\sqrt{s}=7$ TeV pp collisions at different multiplicity classes and the corresponding fitted Tsallis\,--\,Pareto distributions are plotted. 

By going towards more central nucleus-nucleus collisions, though the overall shape of the fitted distribution describes the experimental data, the Data/Theory plot shows an increasing 'oscillation'~\cite{Wilk:2015pva, Rybczynski:2014ura}. This is a known effect and needs further investigation, and it also might be an indication that for future studies, especially in central heavy ion collisions a more sophisticated {\it soft+hard} model might be more appropriate~\cite{artic:gbiro, Shen:2017pyo, Yin:2017pzp}. However, studying of this oscillation effect goes beyond the scope of this paper.

\subsection{Unfolding minimum bias parameters}
\label{subsec:Tq}

In our earlier study we have established a connection between the Tsallis\,--\,Pareto parameters fitted to several spectra of various identified hadrons stemming from minimum bias pp collisions~\cite{BG:entr17}. We have found that both $T$ and $q$ show a tendency increasing with the CM energy in a form of \eref{eq:TCM} and \eref{eq:qCM}. For the sake of consistency, we repeated the fits with a slightly different fitting function \eref{eq:TS} including further CMS experimental results~\cite{Khachatryan:2010nk}. The $T_i$ and $q_i$ parameters are summarized in \tref{tab:new} for the given hadron species.

\begin{table}[H]
\caption{The fitted parameters of \eref{eq:TCM} and \eref{eq:qCM} for minimum bias proton-proton collisions in a center-of-mass energy range of $\sqrt{s}\in[62.7; 7000]$ GeV}
  \label{tab:new}
\begin{tabular}{cccccc}
\br
\textbf{Hadron} & \boldmath{$m$} (GeV) & \boldmath{$q_{1}$} & \boldmath{$q_{2}$} & \boldmath{$T_{1}$} (GeV) & \boldmath{$T_{2}$} (GeV) \\
\mr
$\pi^{\pm}$ & 0.140 & 1.001$\pm$0.005 & 0.016$\pm$0.004 & 0.038$\pm$0.017 & 0.005$\pm$0.002 \\
$K^{\pm} $ & 0.493 & 1.008$\pm$0.024 & 0.019$\pm$0.005 & 0.013$\pm$0.016 & 0.017$\pm$0.002 \\
$p(\bar{p})$ & 0.938 & 1.001$\pm$0.044 & 0.018$\pm$0.007 & 0.016$\pm$0.011 & 0.023$\pm$0.002 \\
\br
\end{tabular}
\end{table}

\begin{figure}[H]
  \centering
  \begin{subfigure}[b]{0.5\textwidth}
    \includegraphics[width=\textwidth]{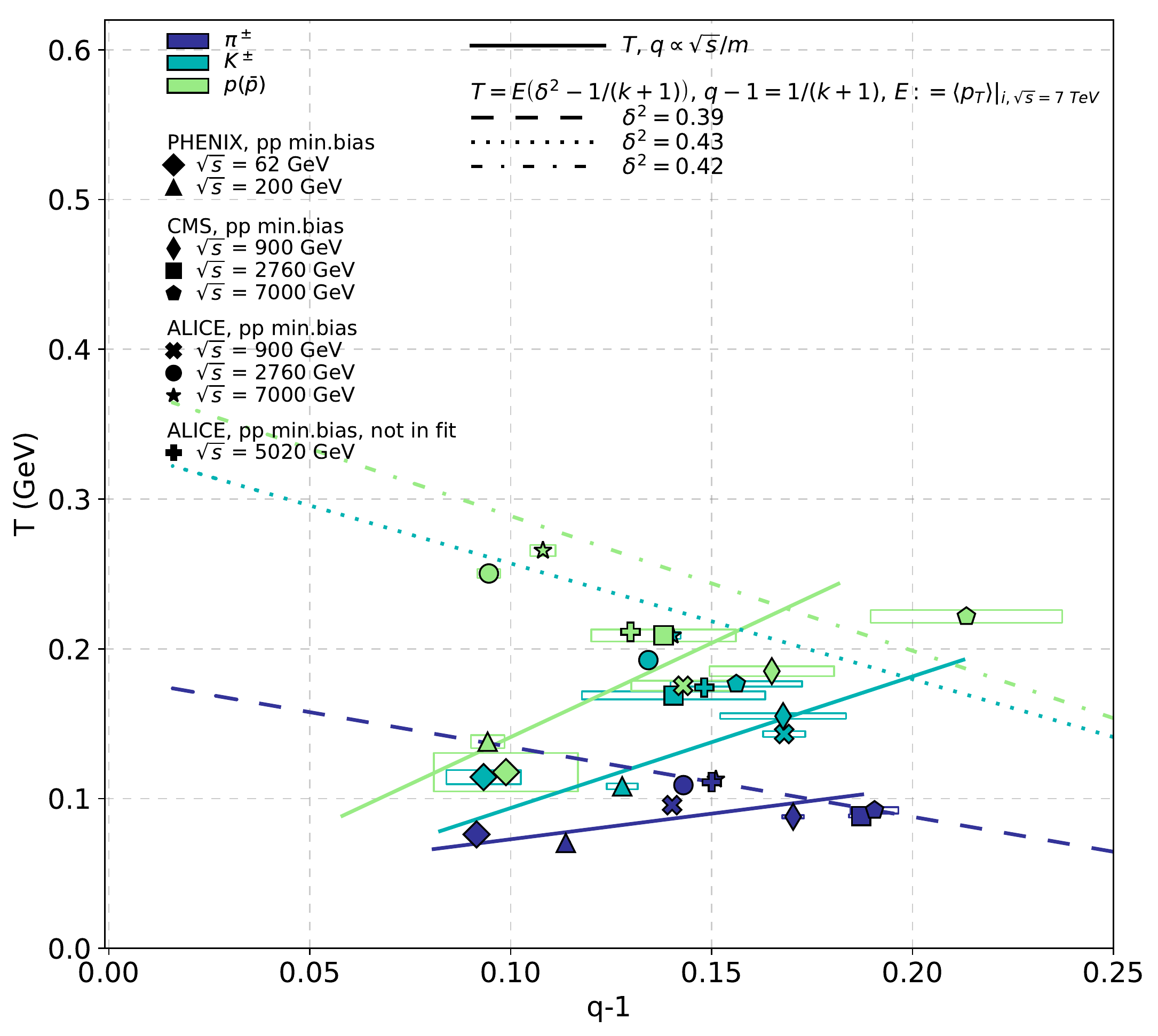}
  \end{subfigure}
  \caption{The fitted parameters $T$ and $q$ of charged pions, kaons and protons stemming from minimum bias pp collisions at various energies. For the legend see the text.}
  \label{fig:Tqfit1}
\end{figure}

The extracted parameters, $T$ and $q$, of charged pions, kaons and protons stemming from minimum bias pp collisions from $\sqrt{s}=$62.4 GeV to 7000 GeV are plotted on \fref{fig:Tqfit1}. The colors are coding the hadron species, while the different markers stand for the different CM energies and experiments. The solid lines are the fits of \eref{eq:TCM} and \eref{eq:qCM}. 

Similarly to our earlier findings, the mass hierarchy (the dependence on the hadron mass) is still strongly manifested~\cite{BG:entr17, BG:maxent16, Biro:2017eip}. In order to test the predicted $\sqrt{s}$ evolution of the $T$ and $q$ parameters, we omitted the $\sqrt{s}=5.02$ TeV data from the fit and plotted it with '+' markers. For the pions, it fits perfectly on the predicted line, while it lies gradually further for the heavier kaons and protons, especially in the $(q-1)$ axis ($\sim 1\%$ and $\sim  2.5$\% respectively). By all means, it is clear that the $q-1$ parameter correlates with the $T$ with a positive slope.

So far only minimum bias experimental data were considered. However, as we have elaborated in \sref{subsec:Tqpars}, there might be a further correlation between $T$ and $q$, which is described by \eref{eq:Tvsqfit} and is related to the multiplicity of the event and it's fluctuations. As it was shown e.g. in \cite{artic:tsbentr}, in case of NB distributed event multiplicity, the $q$ is directly connected to the $k$ parameter of the NBD by $q-1=1/(k+1)$. If as a first order approximation we assume that the $E$ parameter of \eref{eq:Tvsqfit} is the average transverse momentum, $\left<p_T\right>$ of the given hadron species at a given CM energy, the value of the corresponding $\delta^2$ parameters can be calculated, which can be used to test the strength of the correlation. On \fref{fig:Tqfit1} the dashed lines are calculated with  $\left<p_T\right>$ values measured by ALICE at $\sqrt{s}=7$ TeV~\cite{Adam:2016bpr}. The corresponding $\delta^2=\Delta n^2/\left<n\right>^2$ values are listed on this plot and in \tref{tab:Tvsqfit1}.

\begin{table}[H]
  {\small
  \caption{The $\delta^2$ parameters of \eref{eq:Tvsqfit} extracted from $\sqrt{s}=$7 TeV pp data.}
  \label{tab:Tvsqfit1}
  \begin{indented}
    \item[]
    \begin{tabular}{@{}cccc}
      \br
  Hadron & $E:=\left<p_T\right>$ (GeV) & $\delta^2$ & $E\delta^2$ (GeV) \\
  \mr
    $\pi^{\pm}$ & 0.466$\pm$0.010 & 0.39$\pm$0.03 & 0.142$\pm$0.011 \\
    $K^{\pm} $ & 0.773$\pm$0.016 & 0.43$\pm$0.05 & 0.332$\pm$0.039  \\
    $p(\bar{p})$ & 0.900$\pm$0.032 & 0.42$\pm$0.08 & 0.378$\pm$0.073 \\
    \br
  \end{tabular}
\end{indented}
}
\end{table}

The dashed lines describe how the Tsallis $T$ and $q$ parameters depend on the multiplicity fluctuations. The first, quite remarkable feature is that the slope of the correlation changes the sign, while varying the multiplicity. It is also worthwhile to note that for the pions and kaons the predicted line is perfectly crossing through the measured 7 TeV points, while for the protons the $\delta^2$ is underestimated by $\sim 5\%$. We will compare these predictions with real multiplicity measurements in \sref{subsec:Tqmult}.

\subsection{Correlations as functions of event multiplicity}
\label{subsec:Tqmult}

Both parameters, $T$ and $q$ show a strong dependency on the intrinsic system and kinematical parameters like the center-of-mass energy, and an additional mass hierarchy was also observed, as we have shown in \cite{BG:entr17, BG:maxent16} and also other authors in \cite{artic:gribov80, Cleymans:2011in, Santos:2014ota,  Deppman:2012qt, Megias:2015fra, Cleymans:2013tna, artic:lilin1, artic:lilin2,  artic:wilk1, artic:wilk2, artic:wilk3, artic:cleymansjphys, artic:cleymansjphys2, artic:cleymansphyslett, Urmossy:2009jf, Biro:2008er, Bhattacharyya:2017hdc, Tripathy:2016hlg, Yin:2017pzp, Parvan:2016rln, Grigoryan:2017gcg, Khuntia:2017ite}. With the current results we extend this observation: we analyzed identified spectra from proton-proton, proton-nucleus and nucleus-nucleus systems with different CM energies and multiplicity classes.

\begin{figure}
  \centering
  \begin{subfigure}[b]{0.49\textwidth}
    \includegraphics[width=\textwidth]{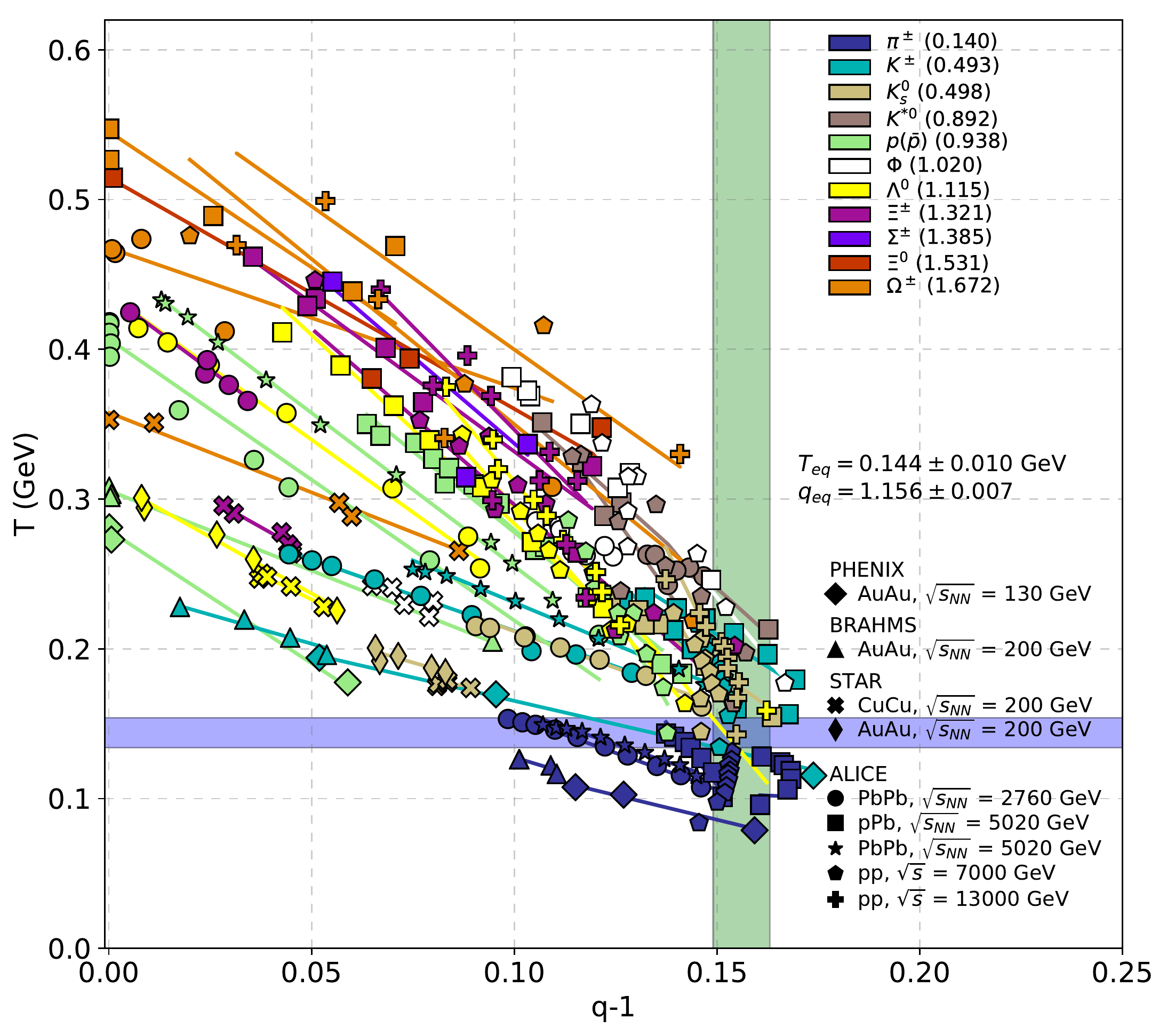}
    \caption{}
    \label{subfig:Tvsq}
  \end{subfigure}
  \begin{subfigure}[b]{0.49\textwidth}
    \includegraphics[width=\textwidth]{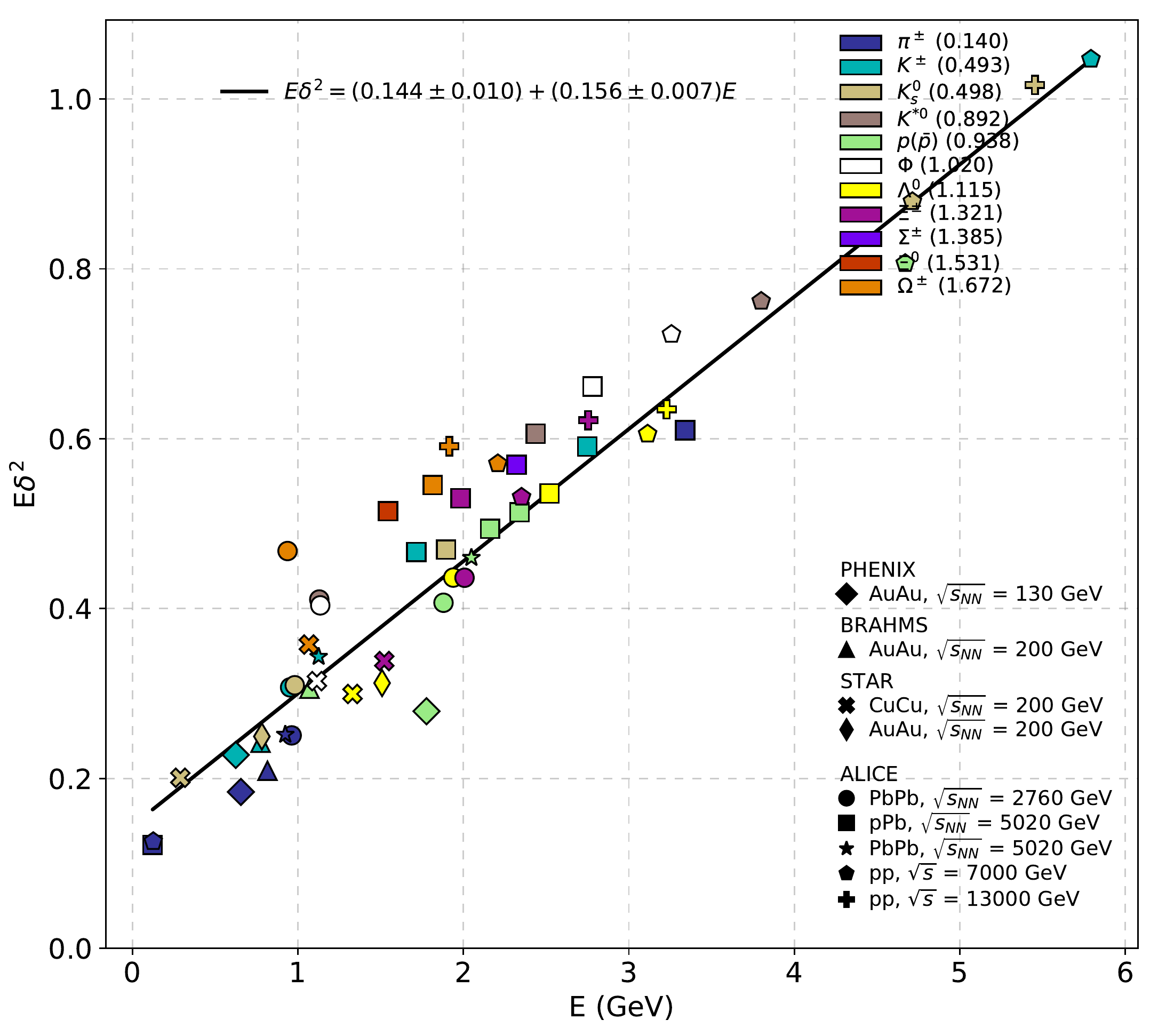}
    \caption{}
    \label{subfig:Evsd2}
    \end{subfigure}
  \caption{\textit{Left panel:} 
  The fitted $T$ and $q$ parameters of all identified hadrons stemming from pp, pA, and AA collisions at various energies and various multiplicity classes. The solid lines are from \eref{eq:Tvsqfit} (each line is at a given CM energy), the parameters are listed in \tref{tab:Tvsqfit2} in \sref{apx:Tvsqfits}. \textit{Right panel:} the fit parameters $E\delta^2$ as a function of $E$.}
  \label{fig:Tvsqfit2}
\end{figure}

The fitted $T$ and $q$ parameters depend strongly on the event multiplicity. In fact, the multiplicity (in relation with system size) has a much larger impact than the CM energy evolution: by going from small to large multiplicity, the change can be as large as of $\mathcal{O}(100\%)$ for the $T$ and of $\mathcal{O}(10\%)$ for the $q$, can be $\mathcal{O}(500\%)$ for $(q-1)$, depending on the given hadron species. As we have concluded in \ref{subsec:Tq}, this correlation can be quantified by the 'Tsallis-thermometer', by fitting \eref{eq:Tvsqfit} on the extracted Tsallis\,--\,Pareto parameters at a given energy. The $T$ and $(q-1)$ parameters for each investigated system are plotted on the left panel in \fref{fig:Tvsqfit2}, the corresponding datasets are listed in the legend and in \tref{tab:exp_sets_2}. The fits of \eref{eq:Tvsqfit} are plotted as solid lines. The fitted $E$ and $\delta^2$ parameters are shown on the right panel and listed in \tref{tab:Tvsqfit2} in \sref{apx:Tvsqfits}.

There are multiple conclusions to be drawn. First of all, on \fref{subfig:Tvsq} there is a very prominent grouping, with all hadrons tending to the $q\approx 1.15$ and $T\approx 0.14$ GeV region on the 'Tsallis-thermometer'. Altogether with this grouping, for the heavier hadrons the multiplicity scaling effect is stronger, covering a larger region of the $T$~-~$(q-1)$ parameter space. Simultaneously, the fitted parameters of \eref{eq:Tvsqfit} are varying as: $\delta^2\sim 0.17-0.38$ and $E\sim 0.1-6.0$ GeV.

An important observation is that these values are not in agreement with those listed in \tref{tab:Tvsqfit1}. 
Although the fitted linear relations of \eref{eq:Tvsqfit} on \fref{subfig:Tvsq} qualitatively matching with the predictions of \fref{fig:Tqfit1} marked with dashed lines there are some important exceptions: at low multiplicity (i.e. in pp and pA collisions) for the lightest pions and kaons there is some kind of '\textit{critical point}': as the multiplicity is increased, there is a value where the correlation changes the sign. For $\pi^\pm$ it is around $\left<\dd N_{ch}/\dd \eta\right>\sim 10-15$, while for the charged and neutral kaons it happens at $\sim 6-10$. For the heavier hadrons like $K^{*0}$ and for $p(\bar{p})$ such a turning point is not visible. However, since the final state in an event consists mostly of pions, this result agrees with the experimental observations, namely that there is a strong "indication that something starts happening around $\dd N_{ch}/\dd \eta\sim$10-20", as stated in~\cite{preghenella:sqm19}.

In order to determine the precise values of the region where the parameters are concentrated, we fit \eref{eq:Tvsqfit} again, but this time the fit parameters are $T$ and $q-1$ (see \fref{subfig:Evsd2}):
\begin{eqnarray}
  E\delta^2&=T_{eq}+(q_{eq}-1)E=(0.144\pm0.010)\textrm{ GeV} + (0.156\pm0.007)E,
\end{eqnarray}
from which the common, grouping parameters (labeled with an $eq$ index) can be read:
\begin{eqnarray}
  T_{eq}&= 0.144\pm0.010\textrm{ GeV}, \label{eq:Tequiv_fit} \\
  q_{eq}&=1.156\pm0.007. \label{eq:qequiv_fit}
\end{eqnarray}
The corresponding regions are plotted as shaded bands on \fref{subfig:Tvsq}. These values are based on the assumption that all hadronic species contribute to the degrees of freedom and we are not in the Boltzmann\,--\,Gibbs limit.

\subsection{Multiplicity scaling}
\label{subsec:mult}

\begin{figure}[H]
  \centering
  \begin{subfigure}[b]{0.495\textwidth}
    \includegraphics[width=\textwidth]{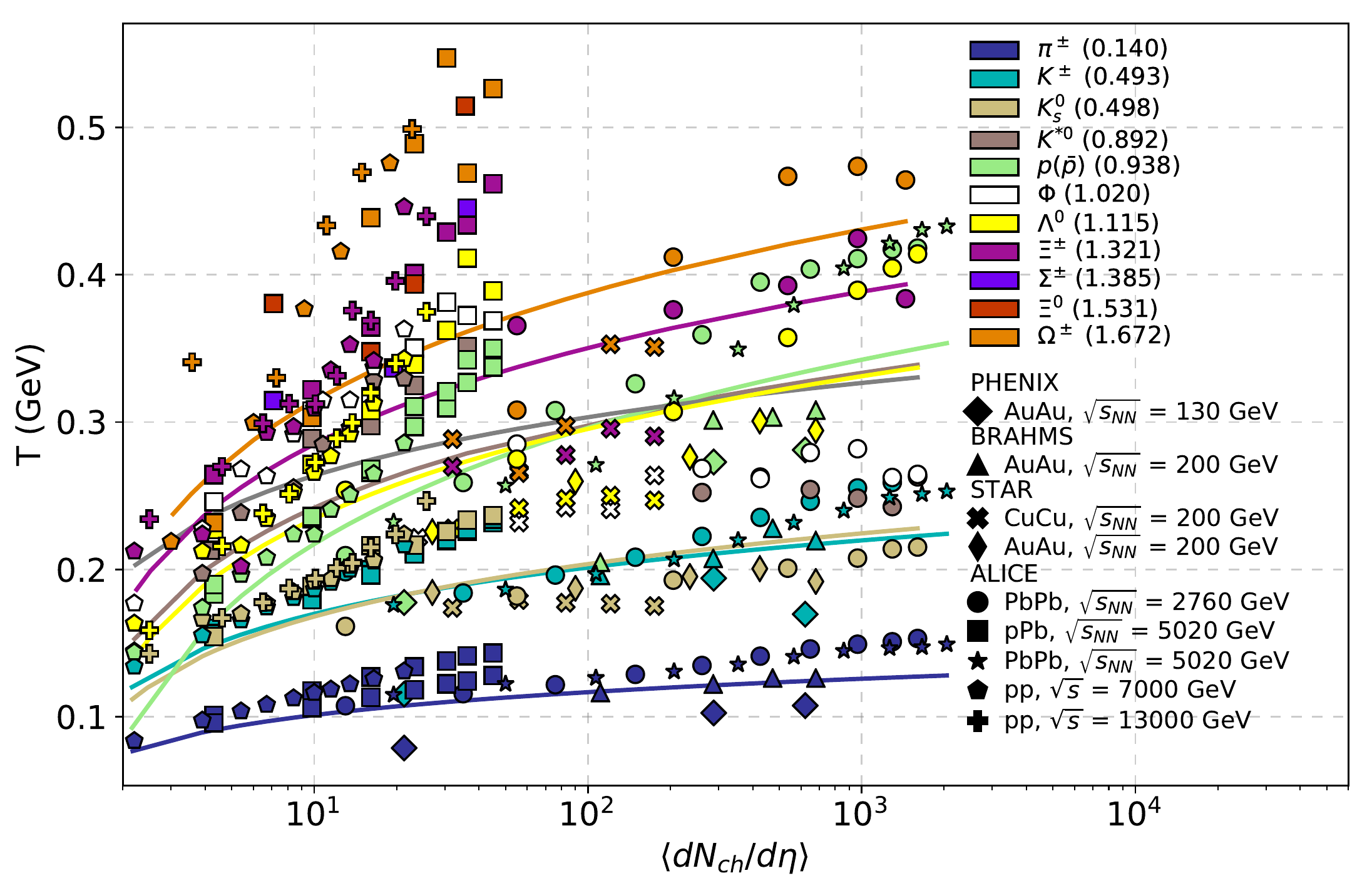}
    \caption{}
    \label{fig:mult_ALL_T}
    \end{subfigure}
  \begin{subfigure}[b]{0.495\textwidth}
    \includegraphics[width=\textwidth]{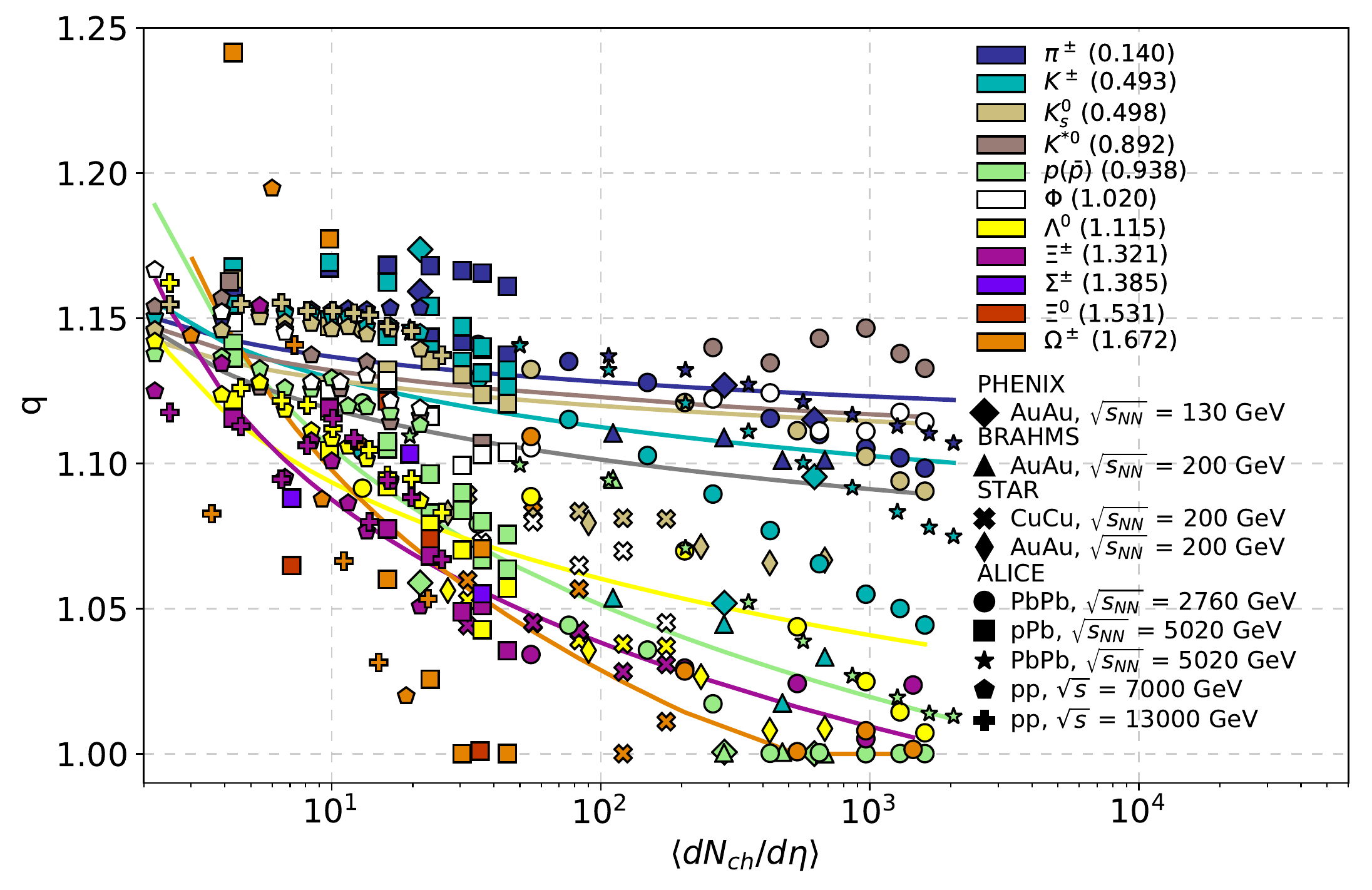}
    \caption{}
    \label{fig:mult_ALL_q}
    \end{subfigure}
  \begin{subfigure}[b]{0.495\textwidth}
    \includegraphics[width=\textwidth]{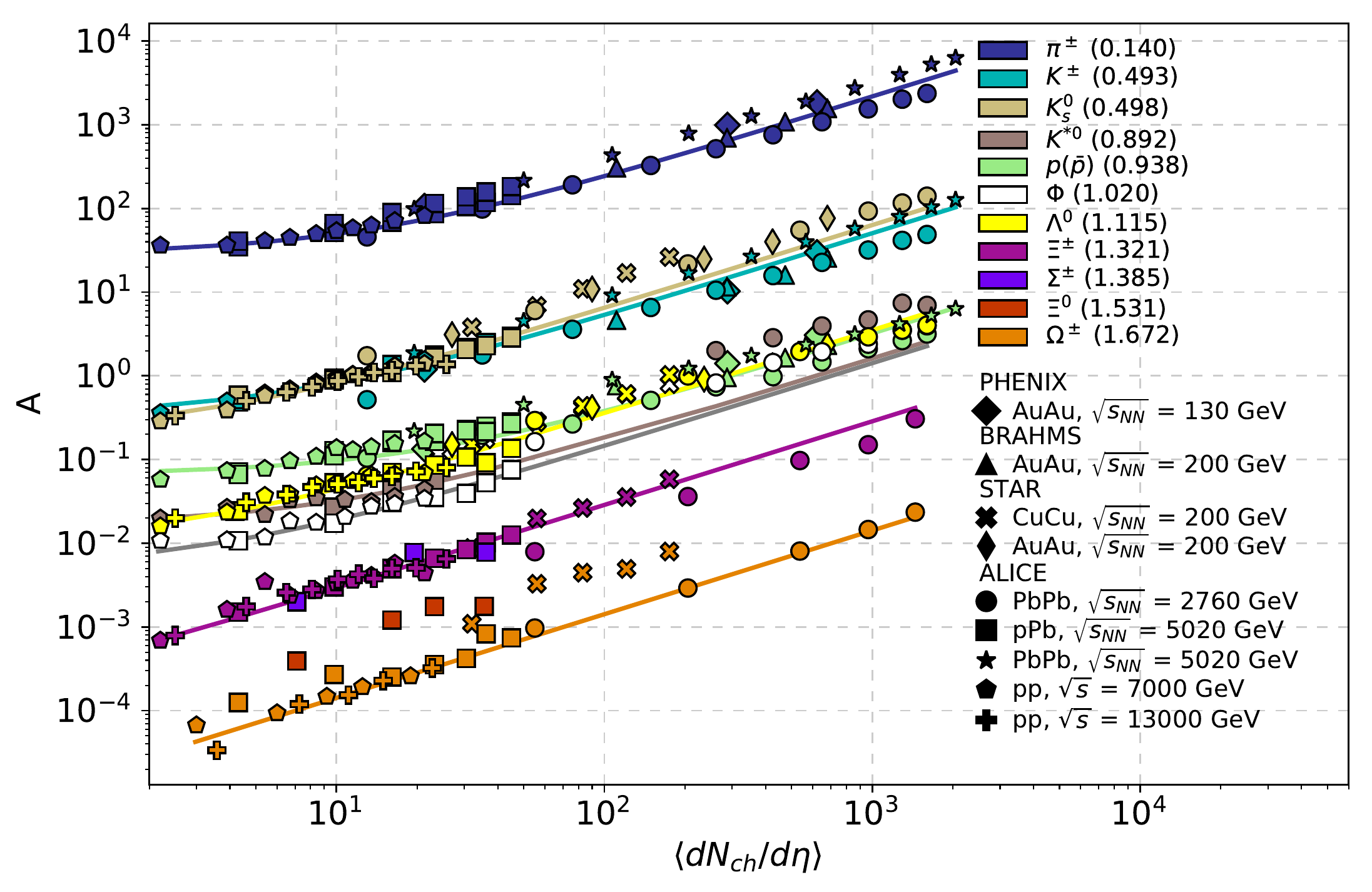}
    \caption{}
    \label{fig:mult_ALL_A}
    \end{subfigure}
  \caption{The fitted $T$, $q$, and $A$ parameters by \eref{eq:parametrization_T}, \eref{eq:parametrization_q} and \eref{eq:parametrization_A} respectively for identified hadrons measured in pp, pA and AA collisions at various collision energies, in the function of the event multiplicity. For the better visibility the errors of the parameters are omitted.}
  \label{fig:mult_ALL}
\end{figure}

On the panels of \fref{fig:mult_ALL} all fitted parameters, $T, q$ and $A$ are plotted as the function of the event multiplicity, $\left<\dd N_{ch}/\dd \eta\right>$. The notations are the same as of on \fref{fig:mult_chi2}: the different markers stand for the collision systems and the hadron species are separated by different colors. For better visibility the errorbars (typically $\mathcal{O}(3-6\%)$ for $q$ and $\mathcal{O}(6-10\%)$) are omitted.

\Fref{fig:mult_ALL_T} shows the fitted $T$ values for all investigated datasets. The system size dependency becomes significantly stronger with the increasing hadron mass: while for the charged pions the increase from the lowest to the highest multiplicity 
means only a moderate 77\% increase in $T$ ($T_{low}=0.084$ GeV and $T_{high}=0.149$ GeV), for protons a much more significant 200\% is observed ($T_{low}=0.144$ GeV and $T_{high}=0.433$ GeV). This indicates the presence of a strong radial flow, which is investigated in detail in the next subsection.

This size effect is not present on \fref{fig:mult_ALL_q}, which shows the fitted $q$ values of all investigated hadron species in the function of the event multiplicity. In contrast to the parameter $T$, here the collision energy plays a more important role in the parameter evolution. At similar multiplicity classes, a given hadron species yields in a higher $q$ value at higher collision energy, which is more expressed at higher multiplicities: for the charged pions at $\left<\dd N_{ch}/\dd \eta\right>=45-50$ the achieved $q$ value is $q\approx1.139$ from the $\sqrt{s_{NN}}=2760$ PbPb data, while $q=1.137-1.161$ from the $\sqrt{s_{NN}}=5020$ pPb data and $q=1.141$ from the PbPb data. On the other hand, at the higher $\left<\dd N_{ch}/\dd \eta\right>=1600-1660$ region it is $q=1.098$ from the $\sqrt{s_{NN}}=2760$ PbPb data and $q=1.110$ from the $\sqrt{s_{NN}}=5020$ PbPb data. It is a slight $\sim0.001\%$ and $\sim1\%$ difference in this two multiplicity regions. However, for the slightly heavier charged kaons the difference already grows to $\sim3.2\%$ in the high-multiplicity region. Although these differences are not huge, identified hadrons are measured with a good statistics and the $q$ values can be extracted with a high precision. 
Consequently, keeping in mind the interpretation of parameter $q$, these results suggest that \textit{system size fluctuations become more significant at higher collision energies}. In agreement with this trend at higher multiplicity collisions the effect is less significant.

The normalizing parameter $A$ (which is usually connected to the interaction volume $V$ in Tsallis\,--\,Pareto distributions) shows a further effect that is presented on \fref{fig:mult_ALL_A}. The mass hierarchy is obviously observed in the normalization factor: the heavier the hadron, the lower the parameter value is - however, the slope is almost the same for all species. Besides, it has the strongest scaling with the average event multiplicity, more or less independently of the size of the colliding nuclei and even the center-of-mass energy.

Both on \fref{fig:mult_ALL_T} and \fref{fig:mult_ALL_q} the aforementioned grouping can be observed, similar to what we have first seen on \fref{fig:Tvsqfit2}. Seemingly all hadrons intersect the ordinate in the region defined by \eref{eq:Tequiv_fit} and \eref{eq:qequiv_fit}: $T_{eq}\approx$0.144 GeV, while $q_{eq}\approx$1.156. The question immediately arises: why are these regions special and what is the connection with the usual thermodynamical models? We will further investigate this question in \sref{sec:thermo}. 

In contrast to the above, for the parameter $A$ on \fref{fig:mult_ALL_A} there is no sign of such grouping. However, the scaling spans through multiple orders of magnitude, which means that $A$ is a normalization factor indeed, and the possible thermal properties are encoded in $T$ and $q$.

As a summary of the fitted parameters, we conclude that the fitted Tsallis\,--\,Pareto parameters are more sensitive at more central/high multiplicity events. For really heavy baryons, however, the $\mu\approx m$ assumption can be questioned. In addition, each parameter is driven in different ways by the collision parameters:
\begin{enumerate}
    \item the parameter $T$ is increasing both with hadron mass and with multiplicity, indicating the presence of the radial flow (mass hierarchy);
    \item the parameter $q$ is decreasing both with hadron mass and with multiplicity;
    \item the parameter $A$ increases with the average event multiplicity.
\end{enumerate}

As it was suggested in sections \ref{subsec:Tq} and \ref{subsec:Tqmult}, the Tsallis\,--\,Pareto parameters depend on the event multiplicity strongly, and therefore the minimum bias values represent a specific value in this dependency. With these observations we can find a parametrization to 'rule them all' and give a prediction for the yield of any hadron species at any collision system. We can use the following parametrization, as an extension of the evolutions \eref{eq:TCM} and \eref{eq:qCM}:
\begin{eqnarray}
    T(\sqrt{s_{NN}}, \left<\dd N_{ch}/\dd \eta\right>, m) & = T_0  + T_1 \ln{\frac{\sqrt{s_{NN}}}{m}}  + T_2 \ln{\ln{\left<\dd N_{ch}/\dd \eta\right>}} , \label{eq:parametrization_T} \\
    q(\sqrt{s_{NN}}, \left<\dd N_{ch}/\dd \eta\right>, m) & = q_0  + q_1 \ln{\frac{\sqrt{s_{NN}}}{m}}  + q_2 \ln{\ln{\left<\dd N_{ch}/\dd \eta\right>}}, \label{eq:parametrization_q} \\
    A(\sqrt{s_{NN}}, \left<\dd N_{ch}/\dd \eta\right>, m) & = A_0  + A_1 \ln{\frac{\sqrt{s_{NN}}}{m}}  + A_2 \left<\dd N_{ch}/\dd \eta\right>, \label{eq:parametrization_A}
\end{eqnarray}
where $T_i$, $q_i$ and $A_i$ ($i=0,1,2$) are the fitting parameters. On the plots of \fref{fig:mult_ALL} the solid lines represent the result of the mass dependent fits of each parameter. The fitted parameters are summarized in \ref{apx:mult_parameters}.

\begin{figure}
  \centering
  \begin{subfigure}[b]{0.49\textwidth}
    \includegraphics[width=\textwidth]{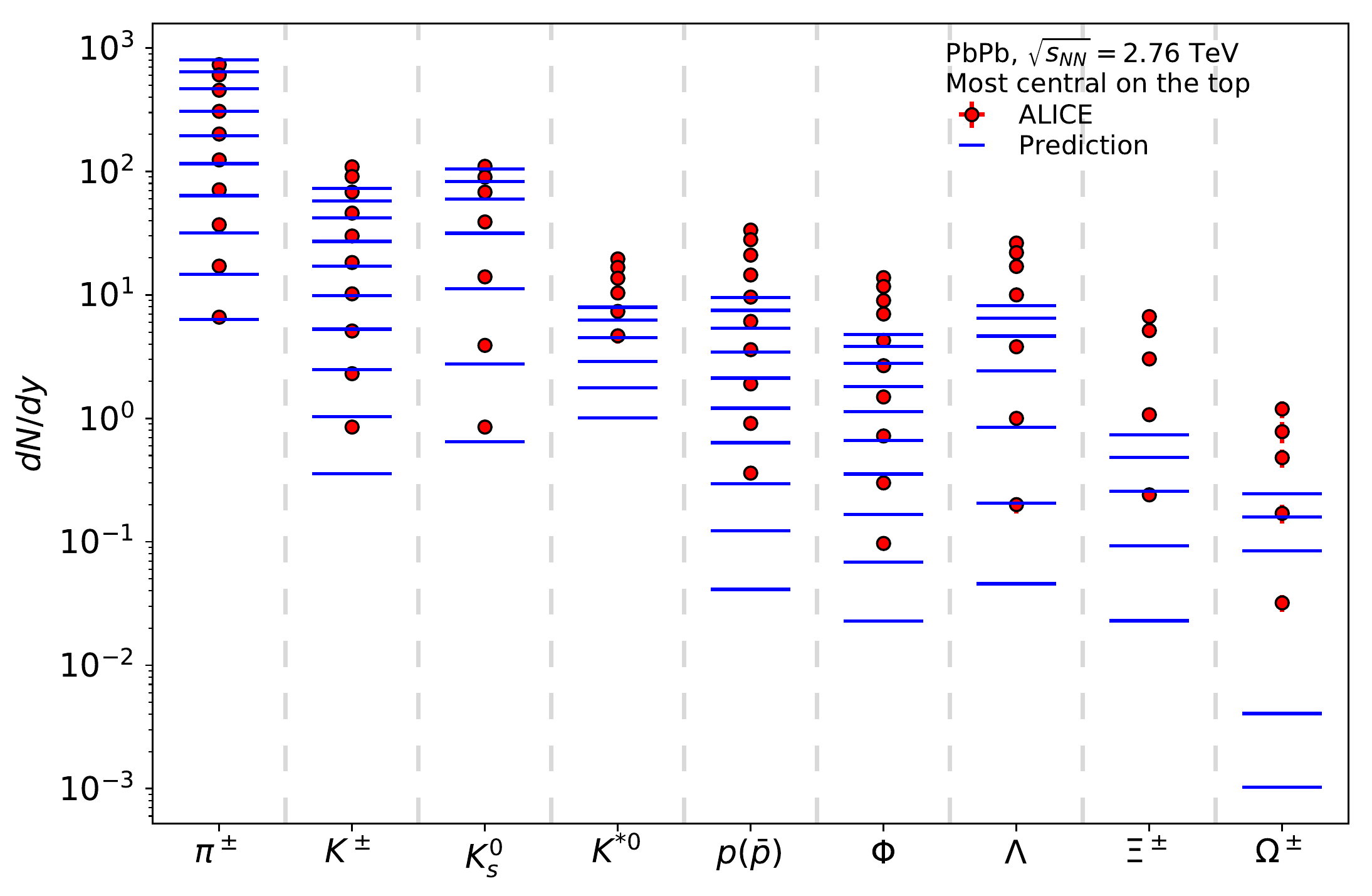}
    \caption{}
    \label{fig:mult_ALL_pred}
  \end{subfigure}  
  \caption{Predictions from \eref{eq:parametrization_T}-\eref{eq:parametrization_A}  (blue lines) for identified hadron yields per unit rapidity at $\sqrt{s_{NN}}=2.76$ TeV center-of-mass energy at PbPb centrality classes, compared with ALICE experimental data (red circles)~\cite{Abelev:2013vea, Abelev:2014laa, Adam:2017zbf, Abelev:2013xaa, ABELEV:2013zaa}. Most central collisions on the top. 
  }
  \label{fig:mult_ALL_pred_nov}
\end{figure}

In \sref{subsec:Tq}, it has been shown experimentally that the event multiplicity is proportional to the center-of-mass energy according to \eref{eq:Nmult}.
Using this formula we can relate the multiplicity variable in the parametrizations above into $\sqrt{s_{NN}}$ and $\left<N_{part}\right>$ dependency. On the other hand, the integration of \eref{eq:TS} over the transverse momentum gives the yield per unit rapidity~\cite{Rath:2019cpe}:
\begin{equation}
    \left.\frac{\dd N}{dy}\right|_{y=0} = 2\pi A T \left[\frac{(2-q)m^2 + 2mT + 2T^2}{(2-q)(3-2q)}\right]\left[1+\frac{q-1}{T}m\right]^{-\frac{1}{q-1}}.
    \label{eq:yieldpred}
\end{equation}
With \eref{eq:yieldpred} and the parametrizations of the Tsallis\,--\,Pareto parameters above, \eref{eq:parametrization_T}-\eref{eq:parametrization_A} we can calculate the yield of a given hadron species at any center-of-mass energy and centrality, using the corresponding number of participant nucleons, $\left<N_{part}\right>$, taken from experimental data. The integrated yields per unit rapidity for various identified hadrons stemming from  $\sqrt{s_{NN}}=2.76$ TeV PbPb collisions with different centrality classes measured by ALICE are shown on \fref{fig:mult_ALL_pred} with red circles, the most central data located on the top~\cite{Abelev:2013vea, Abelev:2014laa, Adam:2017zbf, Abelev:2013xaa, ABELEV:2013zaa}. The predicted yields are plotted with blue lines. 

This na\"ive parametrization gives a fair approximation of the measured data for light hadrons, especially for pions, where the agreement with experimental data is within a few \% for every centrality class. Going towards heavier hadrons and more peripheral collisions the predictions severely underestimate the experimental yield.
  
It is important to emphasize that all the investigations presented so far, were impossible with the usual Boltzmann\,--\,Gibbs statistics. On one hand, for most of the investigated datasets the $p_T$ range of the spectra is too high for a good Boltzmann fit. As the $q\rightarrow 1$ limit would mean that the results of the non-extensive statistics are equivalent to those from the Boltzmann\,--\,Gibbs statistics, from \fref{fig:mult_ALL_q} we can see that this is not the case. It is a unique feature of the non-extensive statistics making us able to investigate such a wide variety of high quality experimental data simultaneously, revealing additional physics encoded in the $q>1$ parameter regime.

\section{Radial flow and kinetic freeze-out temperature}
\label{sec:flow}

The temperature parameter $T$ depends on the mass of the hadron, as is seen on \fref{fig:mult_ALL_T}. This can be interpreted as the presence of radial flow, which increases the transverse momentum of the hadrons proportionally to their mass, therefore resulting in different temperatures instead of the originally common freeze-out temperature~\cite{Adare:2011vy, Schnedermann:1993ws, Csorgo:2001xm}. There are recent studies investigating the radial flow with various models~\cite{Wei:2015oha}, but in this study we apply the formalism that was introduced in \cite{Adare:2011vy}. In this picture the radial flow is described by the following function:
\begin{eqnarray}
  T = T_{fro}+m\left< u_t \right>^2,
  \label{eq:flow1}
\end{eqnarray}
where $T_{fro}$ is the hadron kinetic freeze-out temperature and $\left< u_t \right>$ is the measure of the strength of the average radial transverse flow, which is connected to the averaged transverse velocity via
\begin{eqnarray}
  \left<v_t\right> = \frac{\left< u_t \right>}{\sqrt{1+\left< u_t \right>^2}}.
  \label{eq:flow2}
\end{eqnarray}
In this model it is also assumed that all the investigated hadrons freeze-out at the same time, at the same temperature value.

We recall here that although the $T$ arising from the non-extensive statistics is a temperature-like parameter, obeying thermodynamical relations, but as we have mentioned in \sref{sec:nonext} it is not necessarily equivalent with the usual Boltzmann\,--\,Gibbs temperature: $T\neq T_{BG}$. Therefore, in principle there is no reason to assume that performing a similar blast wave fit with the non-extensive $T$ would result in any comparable radial flow values. Especially because most of the $q$ values are definitely not equal to 1, which would mean the equivalence with the Boltzmann\,--\,Gibbs statistics. Our aim in this section is to perform the freeze-out and $\left<v_t\right>$ fits and compare the non-extensive theory with more conventional models.
  
\begin{figure}[H]
  \centering
  \begin{subfigure}[b]{0.49\textwidth}
    \includegraphics[width=\textwidth]{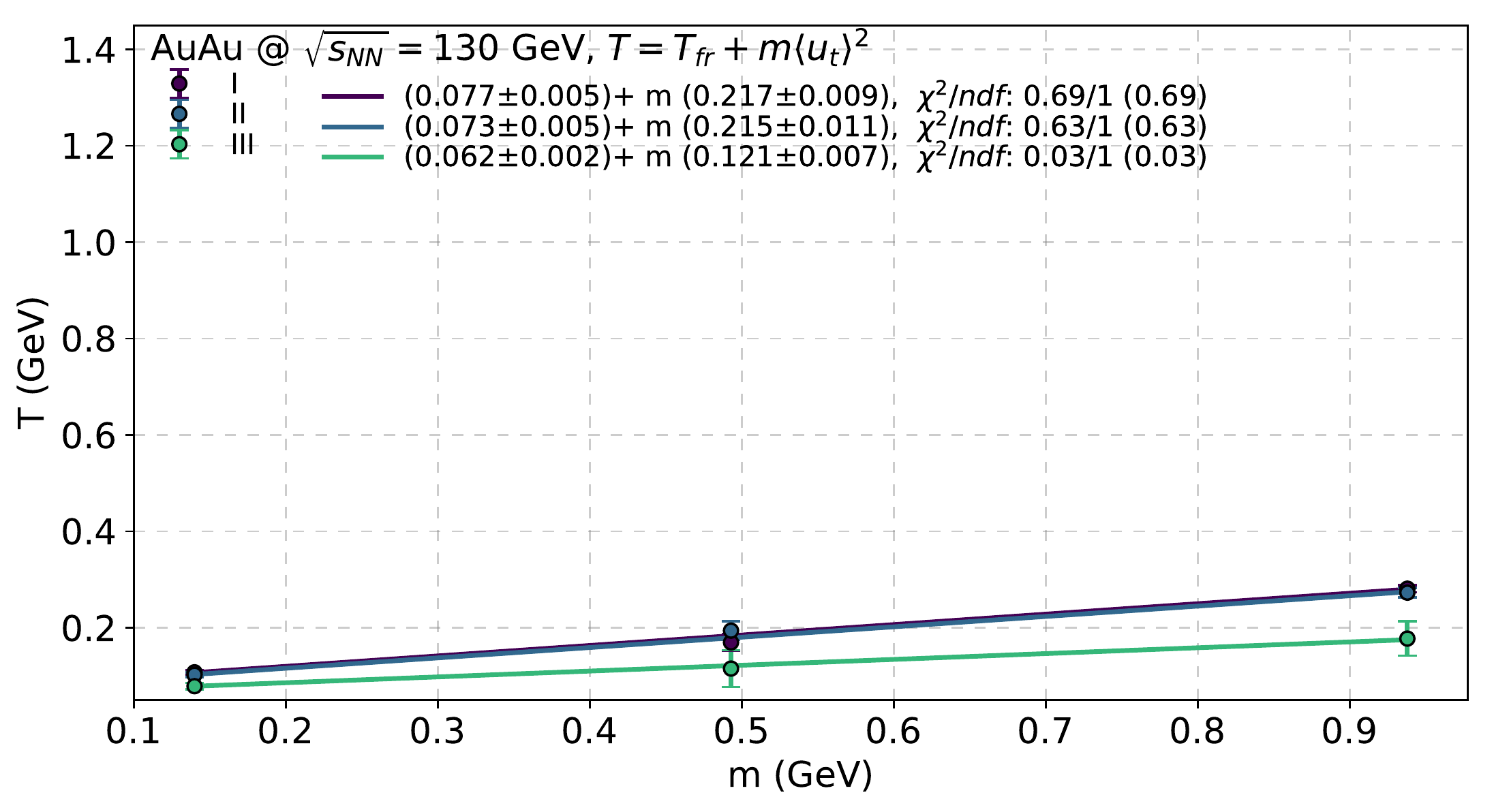}
    \caption{}
    \end{subfigure}
    \begin{subfigure}[b]{0.49\textwidth}
  \includegraphics[width=\textwidth]{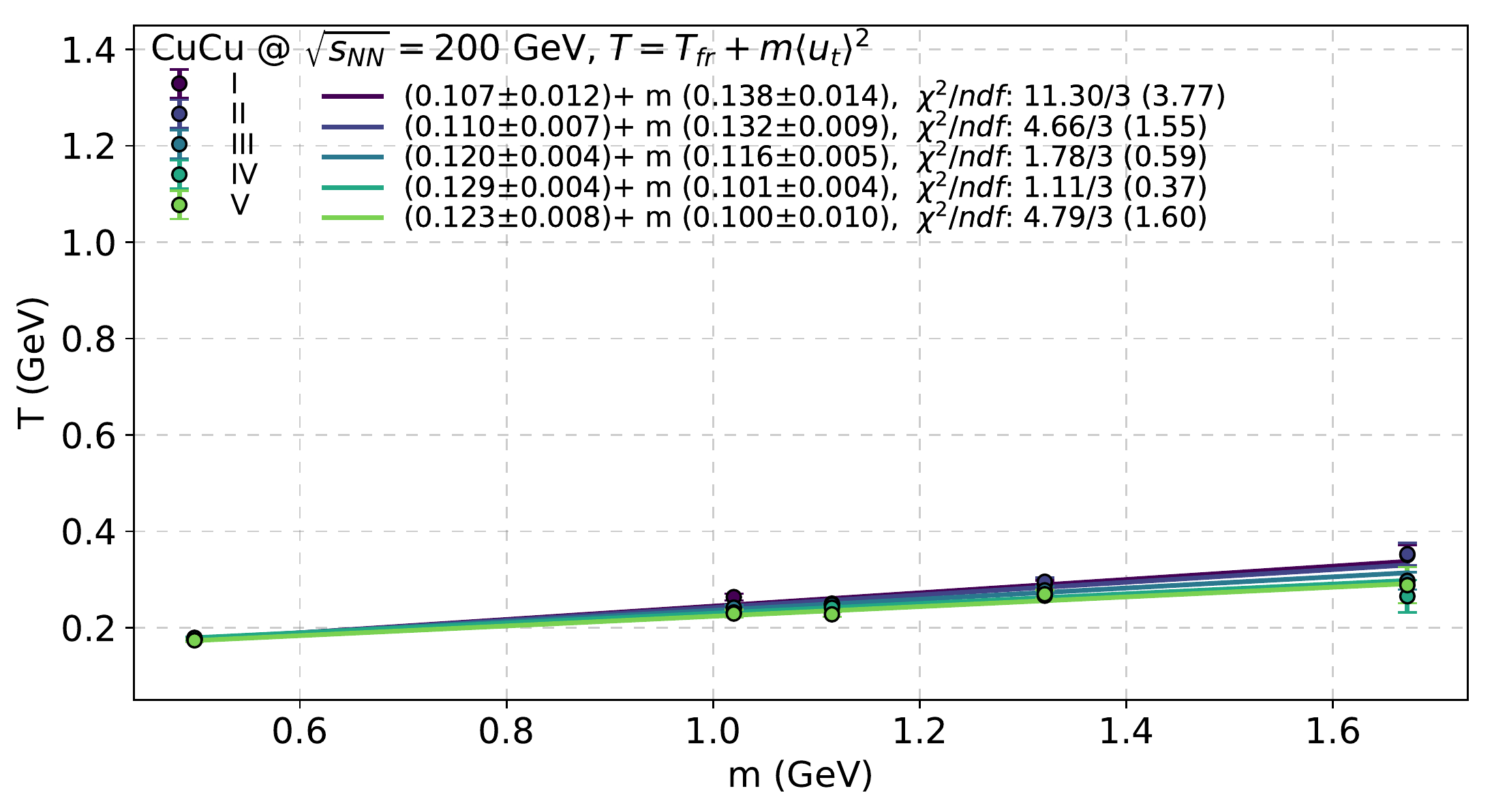}
    \caption{}
  \end{subfigure}
  \begin{subfigure}[b]{0.49\textwidth}
    \includegraphics[width=\textwidth]{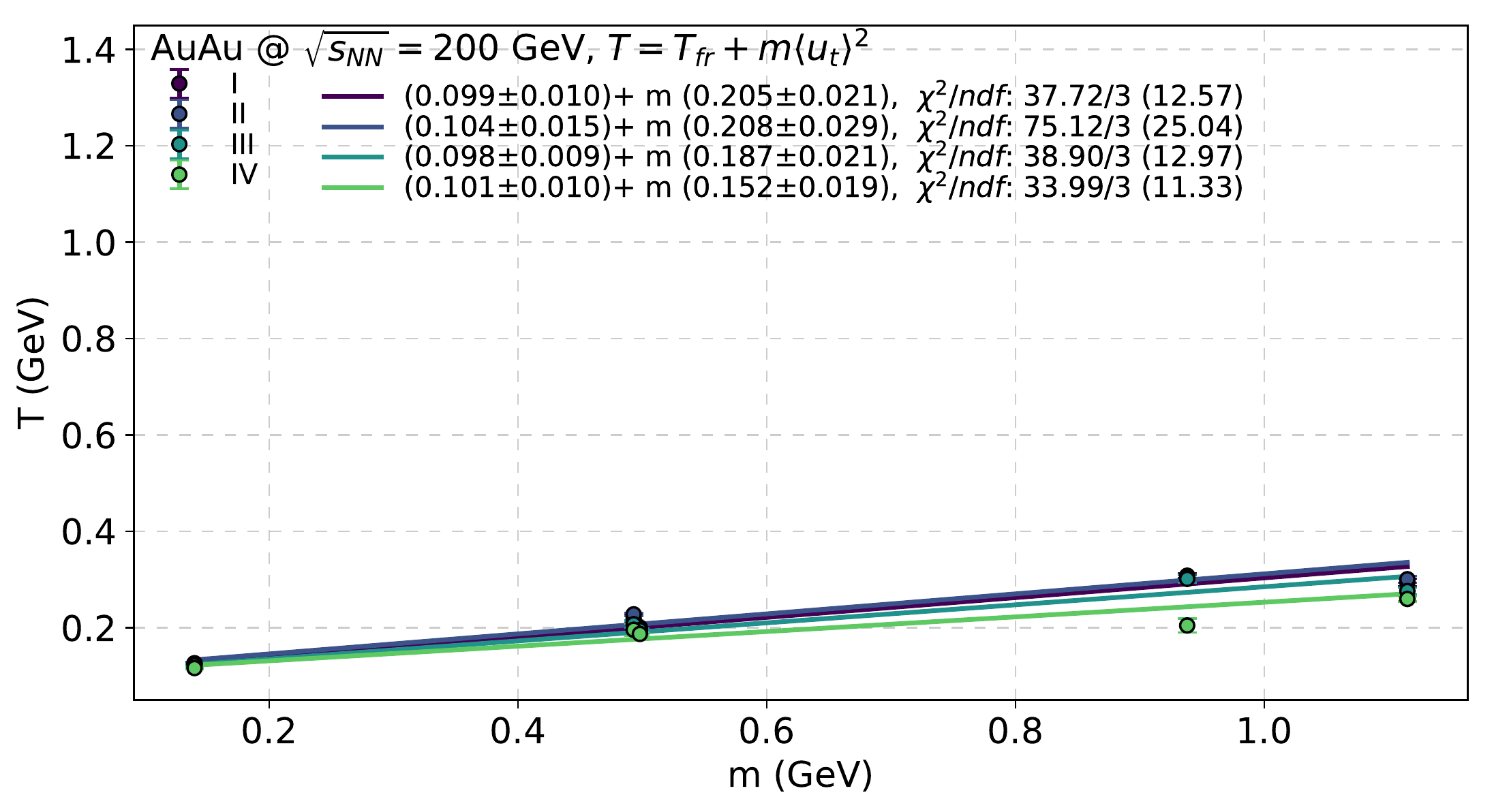}
    \caption{}
    \end{subfigure}
    \begin{subfigure}[b]{0.49\textwidth}
    \includegraphics[width=\textwidth]{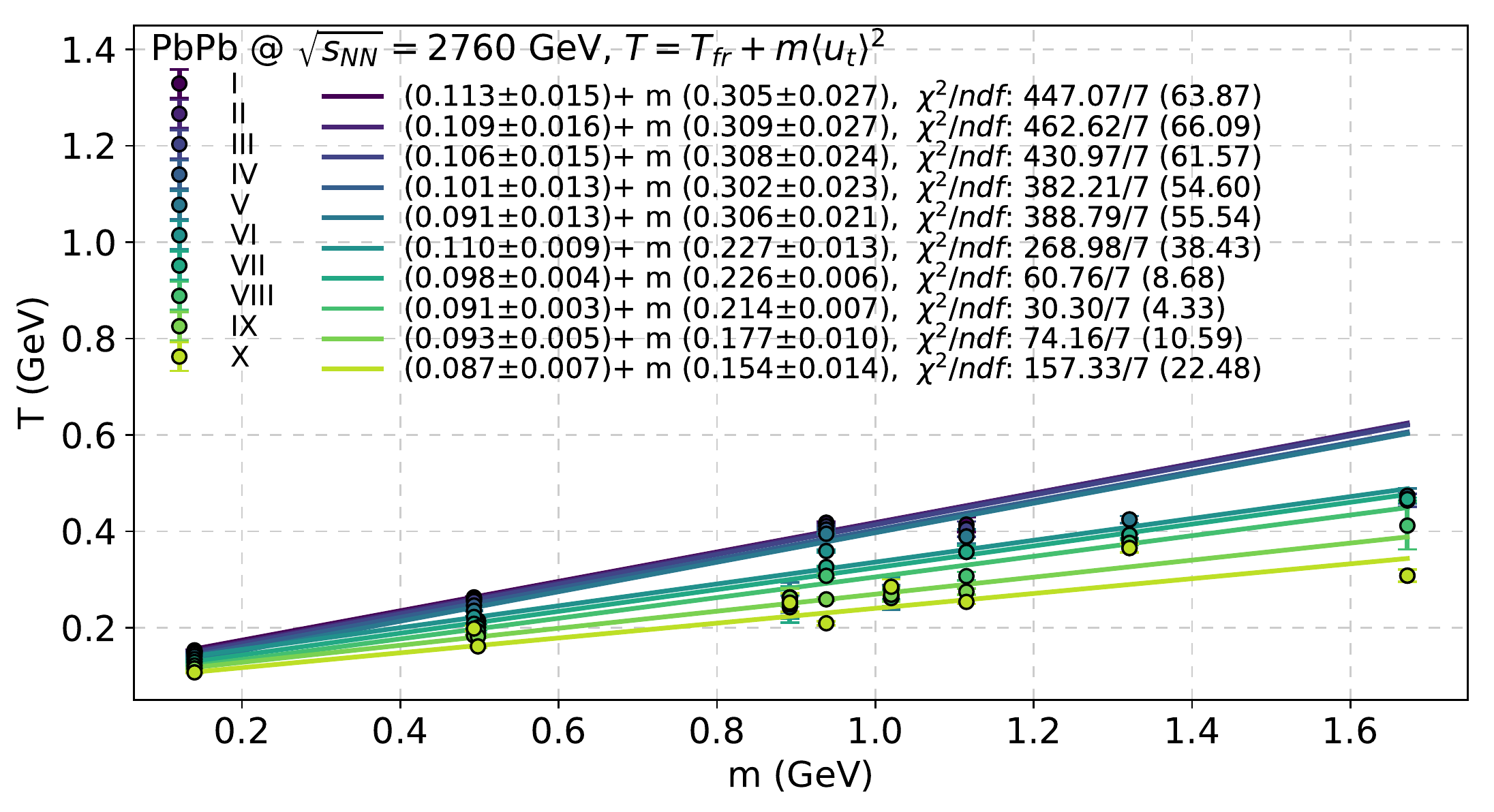}
    \caption{}
    \end{subfigure}
    \begin{subfigure}[b]{0.49\textwidth}
    \includegraphics[width=\textwidth]{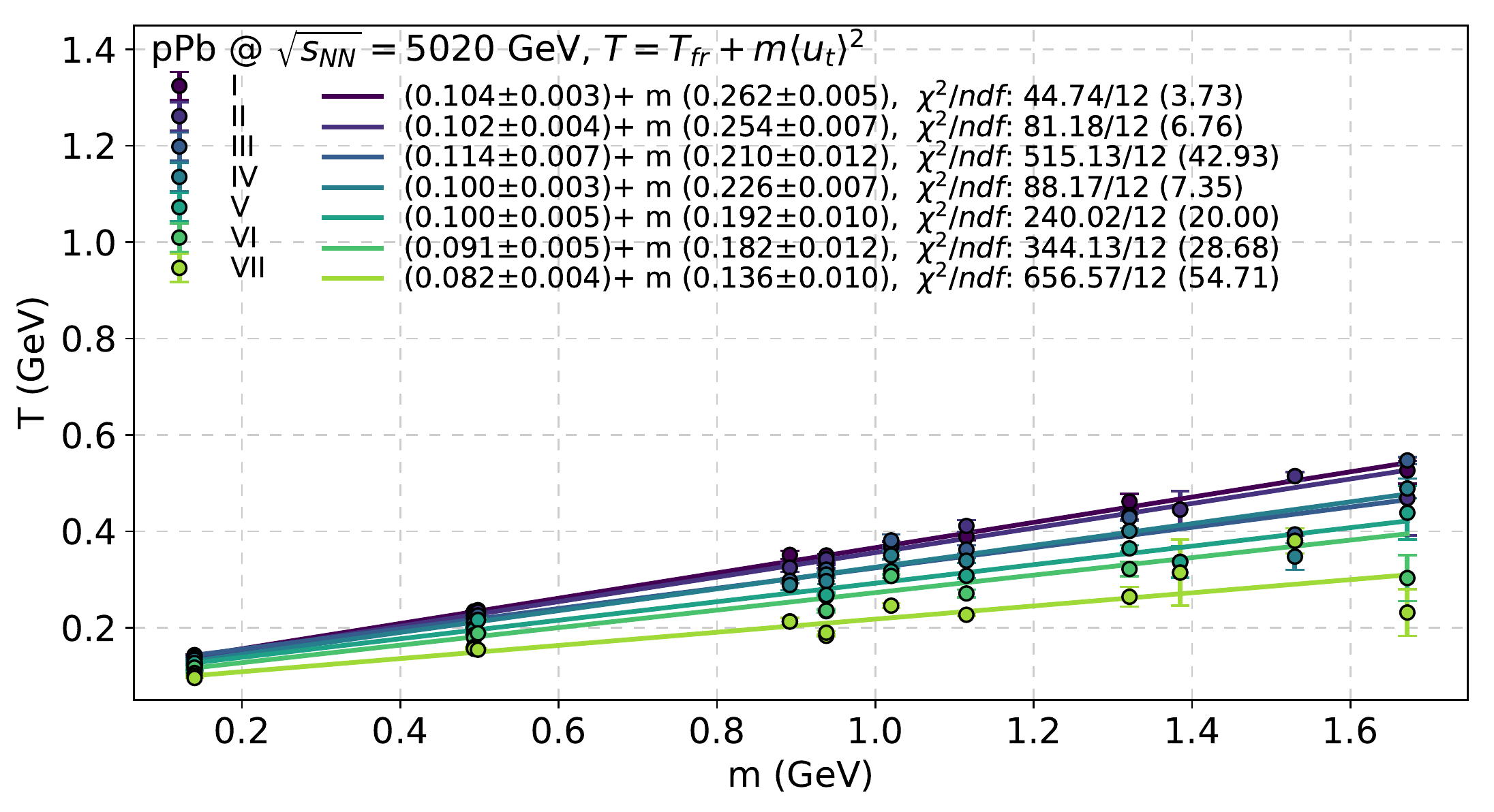}
    \caption{}
    \end{subfigure}
  \begin{subfigure}[b]{0.49\textwidth}
    \includegraphics[width=\textwidth]{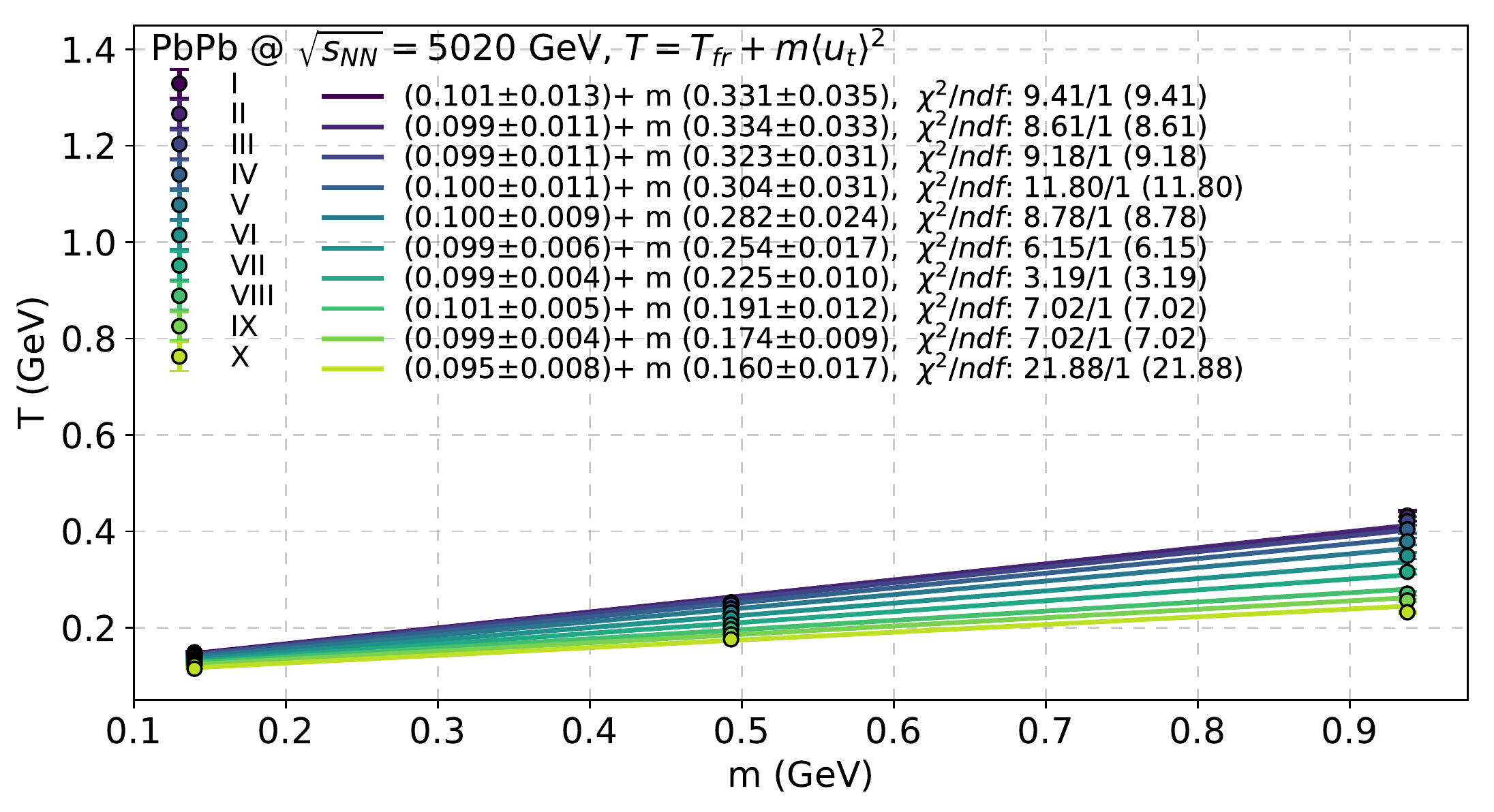}
    \caption{}
    \end{subfigure}
    \begin{subfigure}[b]{0.49\textwidth}
  \includegraphics[width=\textwidth]{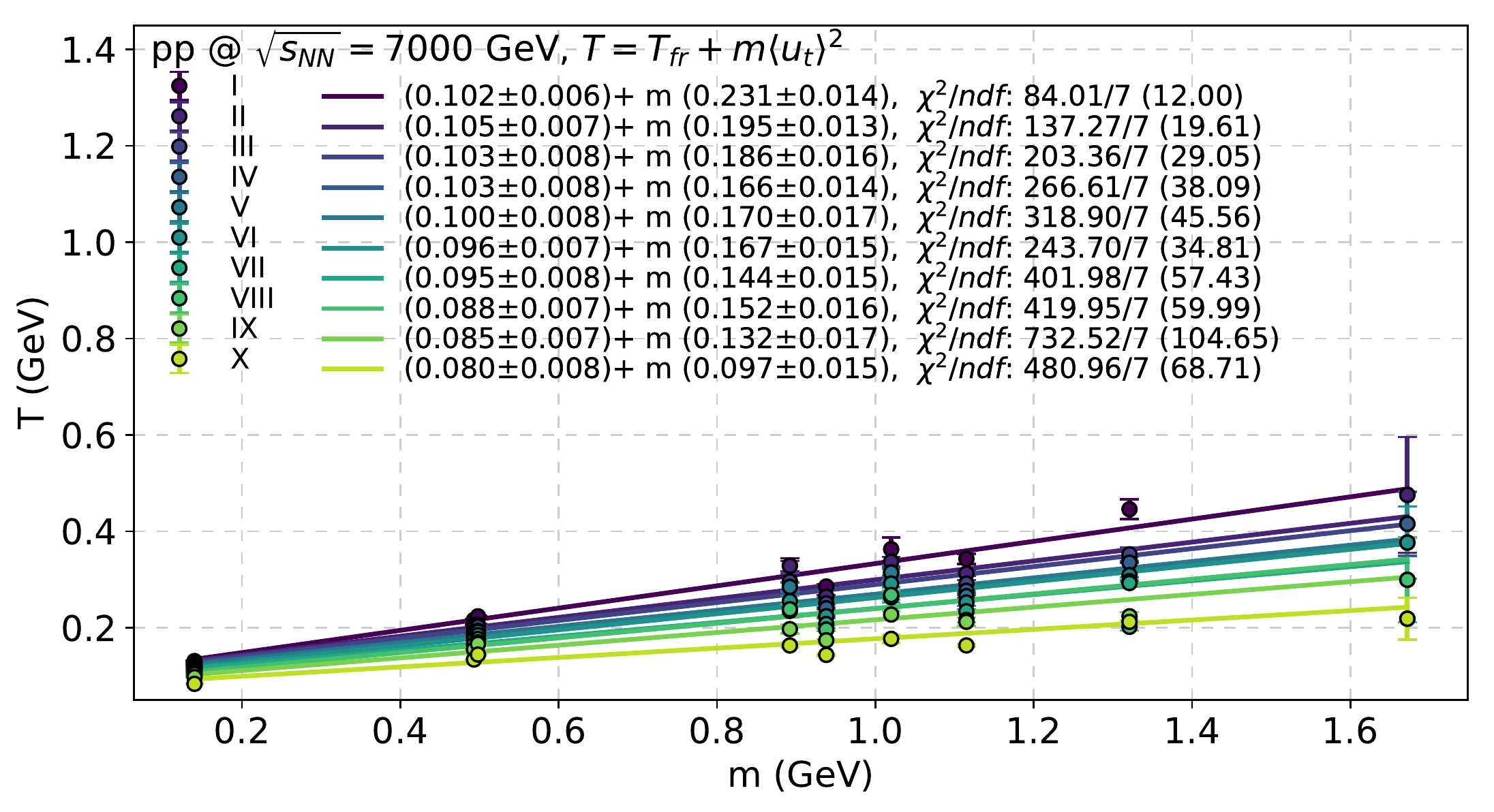}
    \caption{}
  \end{subfigure}
  \begin{subfigure}[b]{0.49\textwidth}
    \includegraphics[width=\textwidth]{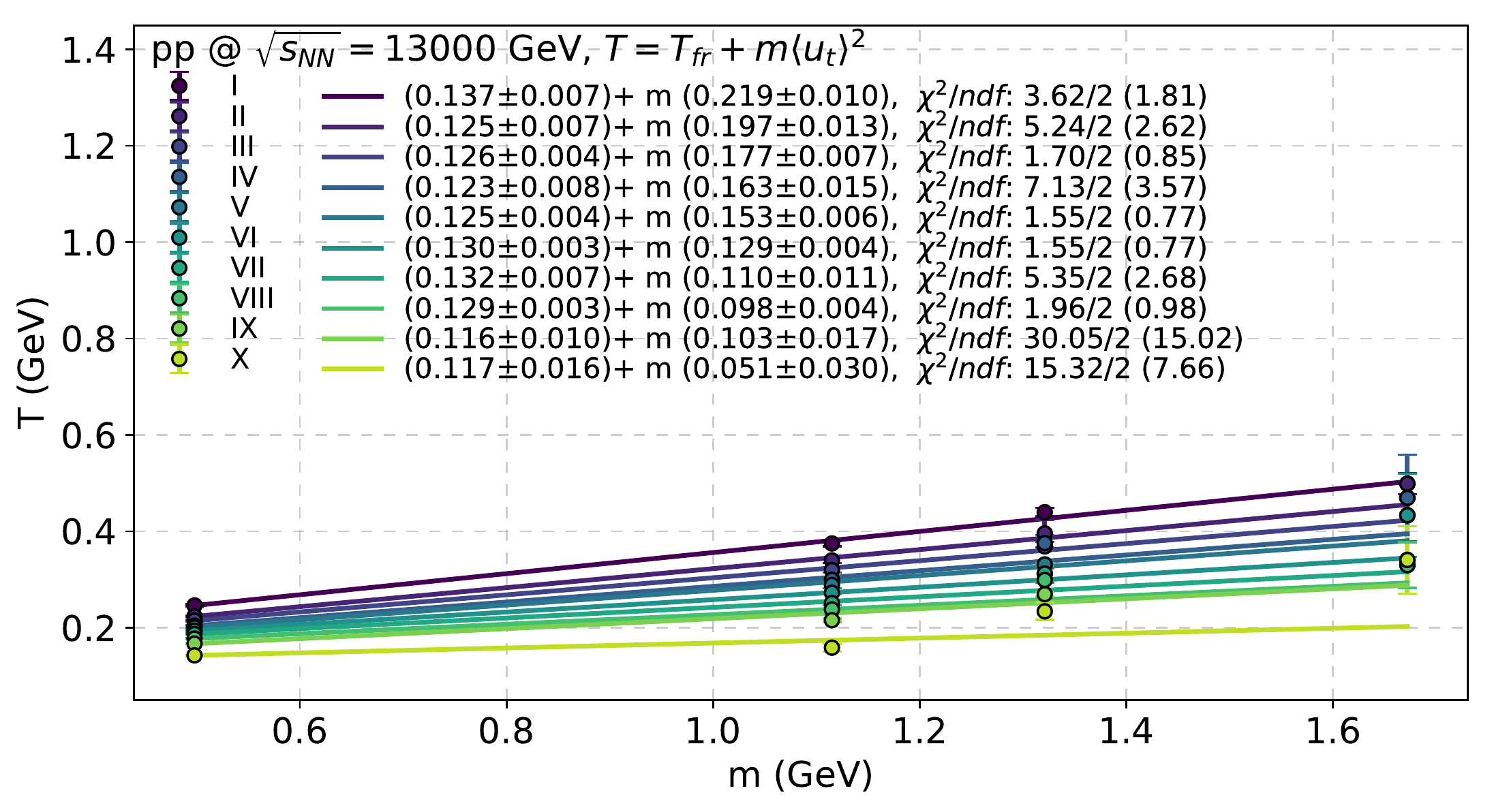}
    \caption{}
    \end{subfigure}
    \caption{Mass dependence of the Tsallis\,--\,Pareto $T$ parameters at various collision systems and multiplicity classes. The solid lines represent the fits of \eref{eq:flow1}.}
    \label{fig:flow_3}
  \end{figure}
  
On the panels of \fref{fig:flow_3} the fitted Tsallis\,--\,Pareto $T$ parameters are plotted in the function of the hadron mass. On each plot the multiplicity classes are indicated by roman numbers, where '\textbf{I}' is the highest multiplicity class and the largest number is the lowest multiplicity class. The fitted functions with the parameter values are indicated by solid lines.

\begin{figure}[H]
  \centering
  \begin{subfigure}[b]{0.49\textwidth}
    \includegraphics[width=\textwidth]{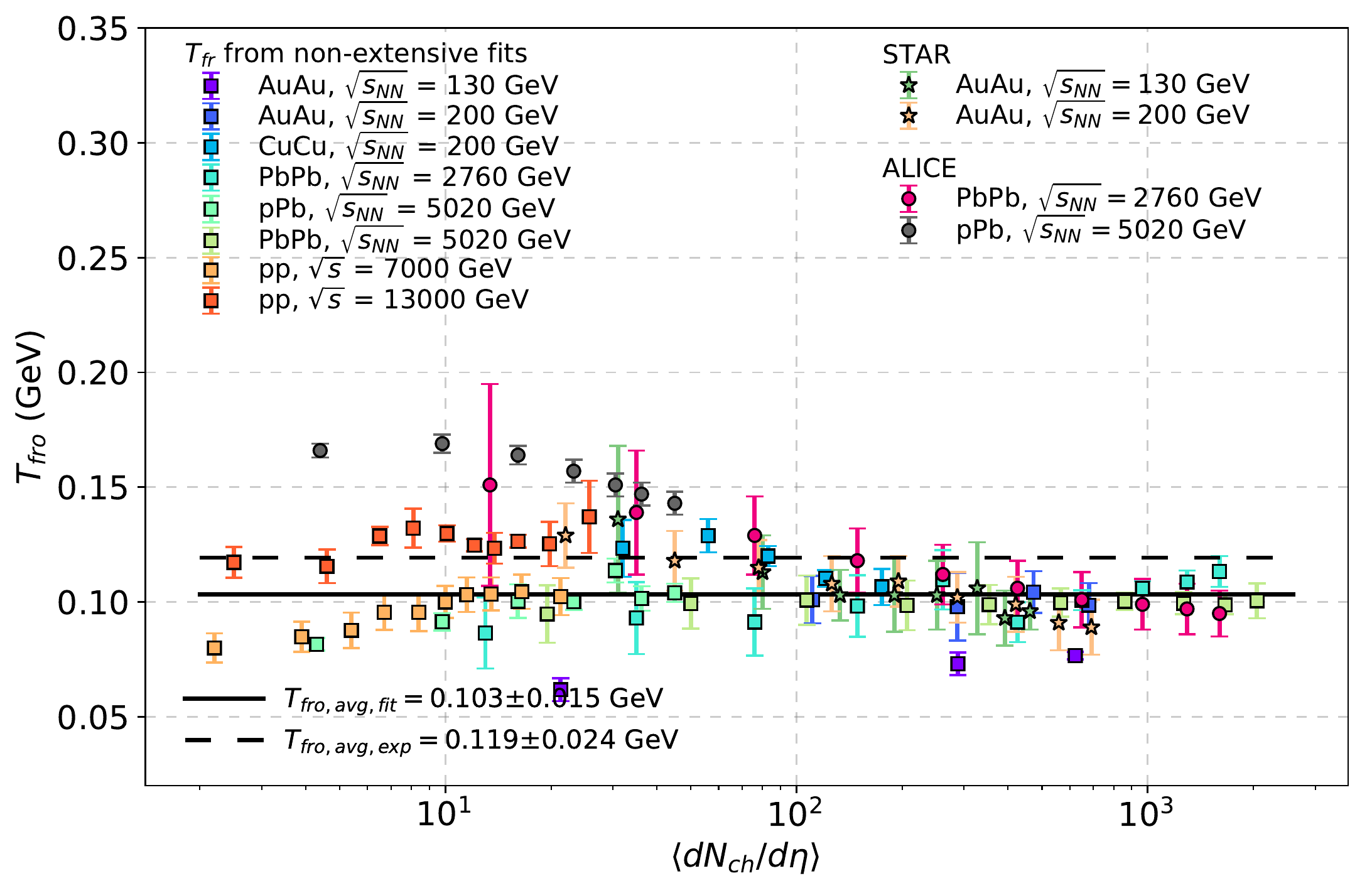}
  \caption{}
    \label{fig:Tfro}
    \end{subfigure}
    \begin{subfigure}[b]{0.49\textwidth}
  \includegraphics[width=\textwidth]{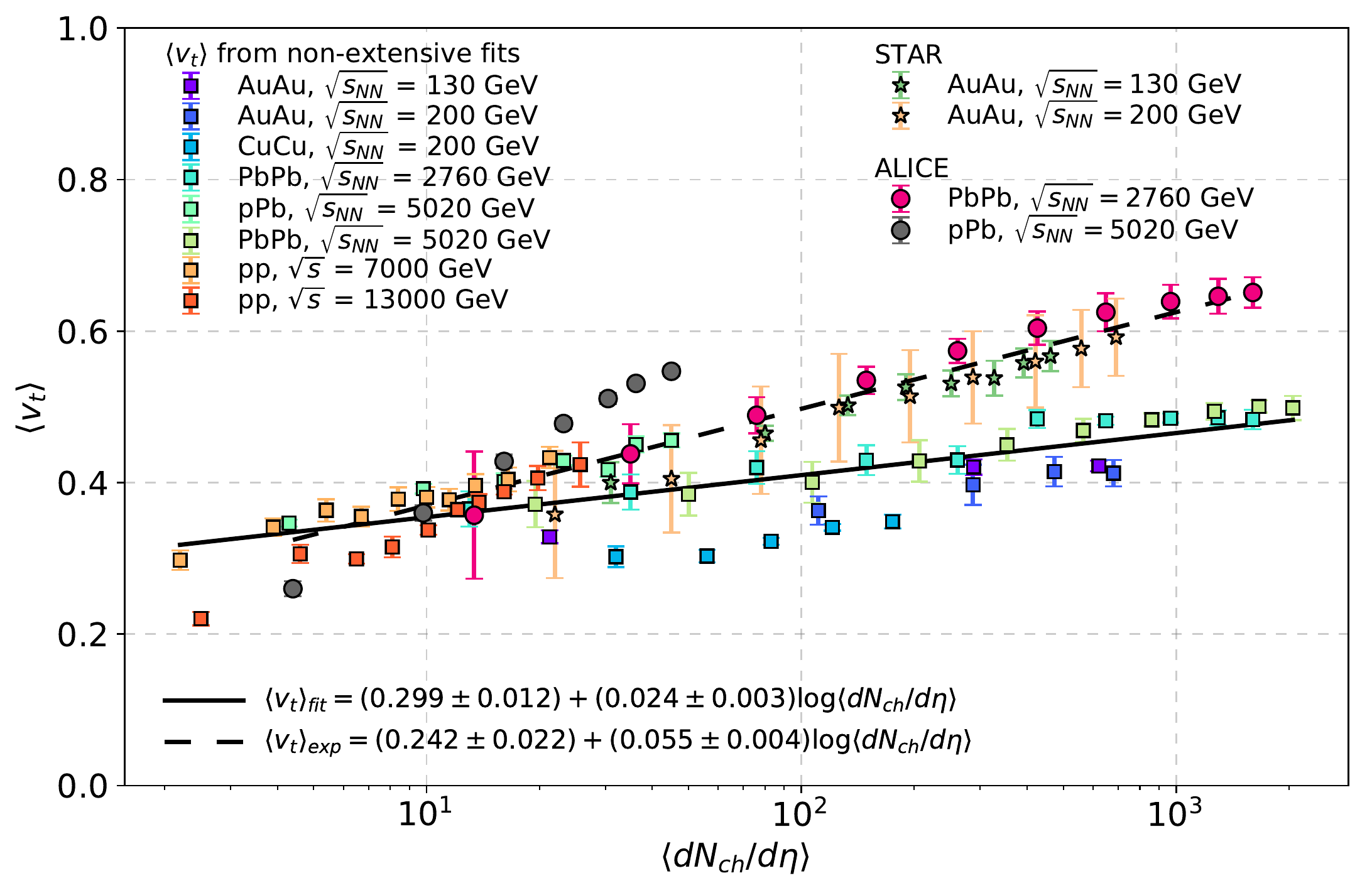}
  \caption{}
  \label{fig:flowvelo}
  \end{subfigure}
  \caption{The freeze-out temperature from \eref{eq:flow1} and the averaged transverse velocity from \eref{eq:flow2} in the function of average event multiplicity, for various collision systems.}
  \label{fig:flow_4}
\end{figure}

  The resulted kinetic freeze-out temperatures, $T_{fro}$ and the averaged transverse velocities in the function of the average event multiplicities are plotted on figures \ref{fig:Tfro} and \ref{fig:flowvelo} respectively, along with STAR and ALICE experimental data~\cite{Abelev:2013haa, Abelev:2013vea, Abelev:2008ab, Adamczyk:2017iwn}. The resulted $T_{fro}$ freeze-out temperature is approximately constant with an average of
  \begin{equation}
    \left<T_{fro}\right>=0.103\pm0.015 \textrm{\ GeV,}
    \label{eq:Tfrofit}
  \end{equation}
varying in a range of $[0.09-0.14]$ GeV. Though our results are compatible with the Beam Energy Scan results, the agreement becomes especially good at higher multiplicity values, i.e. at $\left<\dd N_{ch}/\dd \eta\right> >100$.

The average transverse radial flow velocity, $\left<v_t\right>$ shows a slightly increasing tendency with the multiplicity, which is in agreement with the blast wave analysis of the data. Linear fits presented on \fref{fig:flowvelo} reveal that the multiplicity dependence of the transverse radial flow velocity is smaller, than the values can be extracted from the non-extensive approach:
  \begin{eqnarray}
    \left<v_t\right> & = (0.299\pm0.012)+(0.024\pm0.003)\ln{\left<\dd N_{ch}/\dd \eta\right>}.
    \label{eq:flowfit}
  \end{eqnarray}
The agreement is better within the lower multiplicity regime. The main reason of this difference is that while in the experimental blast wave studies only light flavor hadrons are considered, here we fitted all hadrons simultaneously, including the heavy (multi)strange hadrons as well. 

\section{Thermodynamical validation within the non-extensive approach}
\label{sec:thermo}

\begin{figure}[H]
  \centering
  \begin{subfigure}[b]{0.494\textwidth}
    \includegraphics[width=\textwidth]{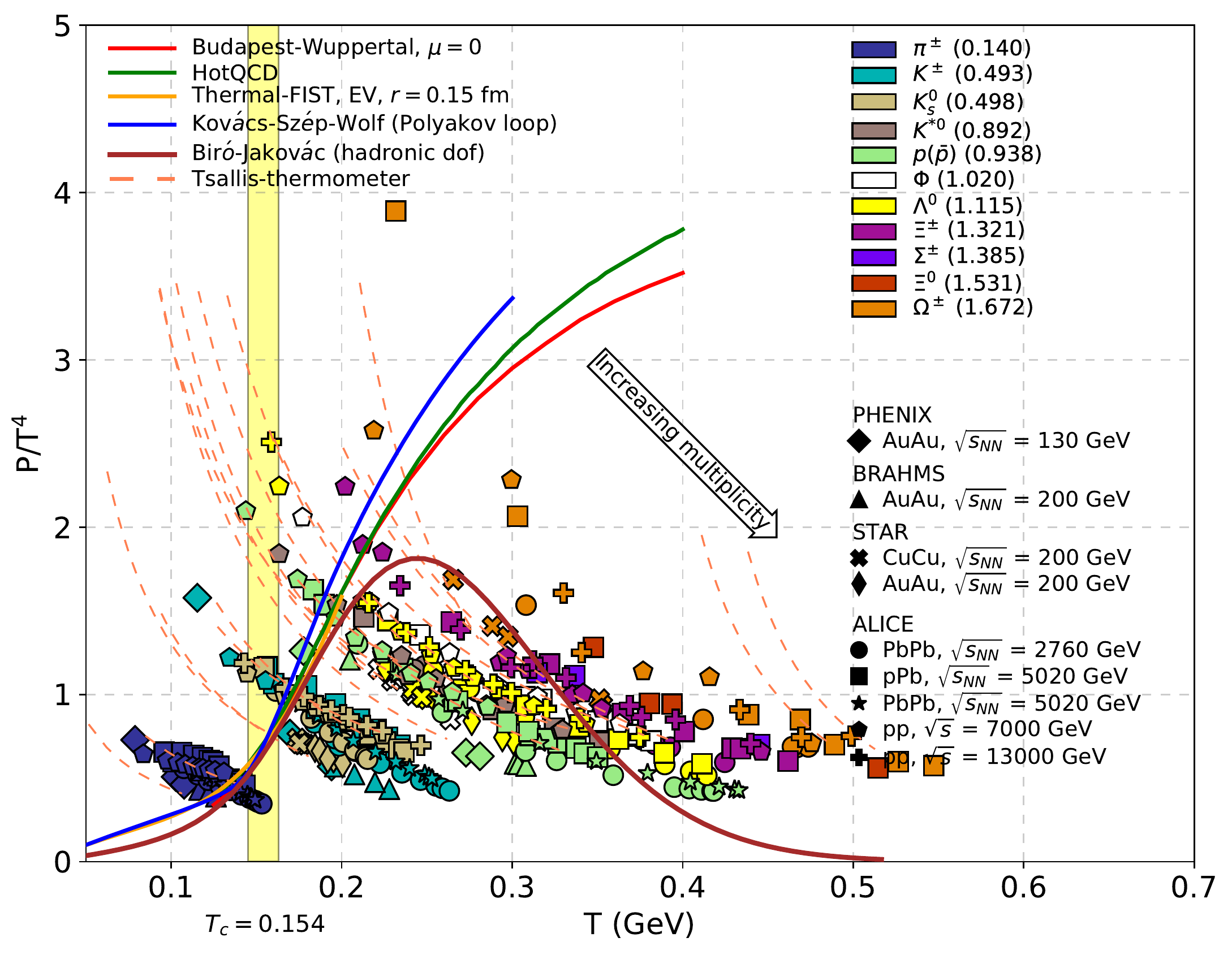}
    \caption{}
    \label{fig:mult_ALL_PT}
    \end{subfigure}
    \begin{subfigure}[b]{0.494\textwidth}
      \includegraphics[width=\textwidth]{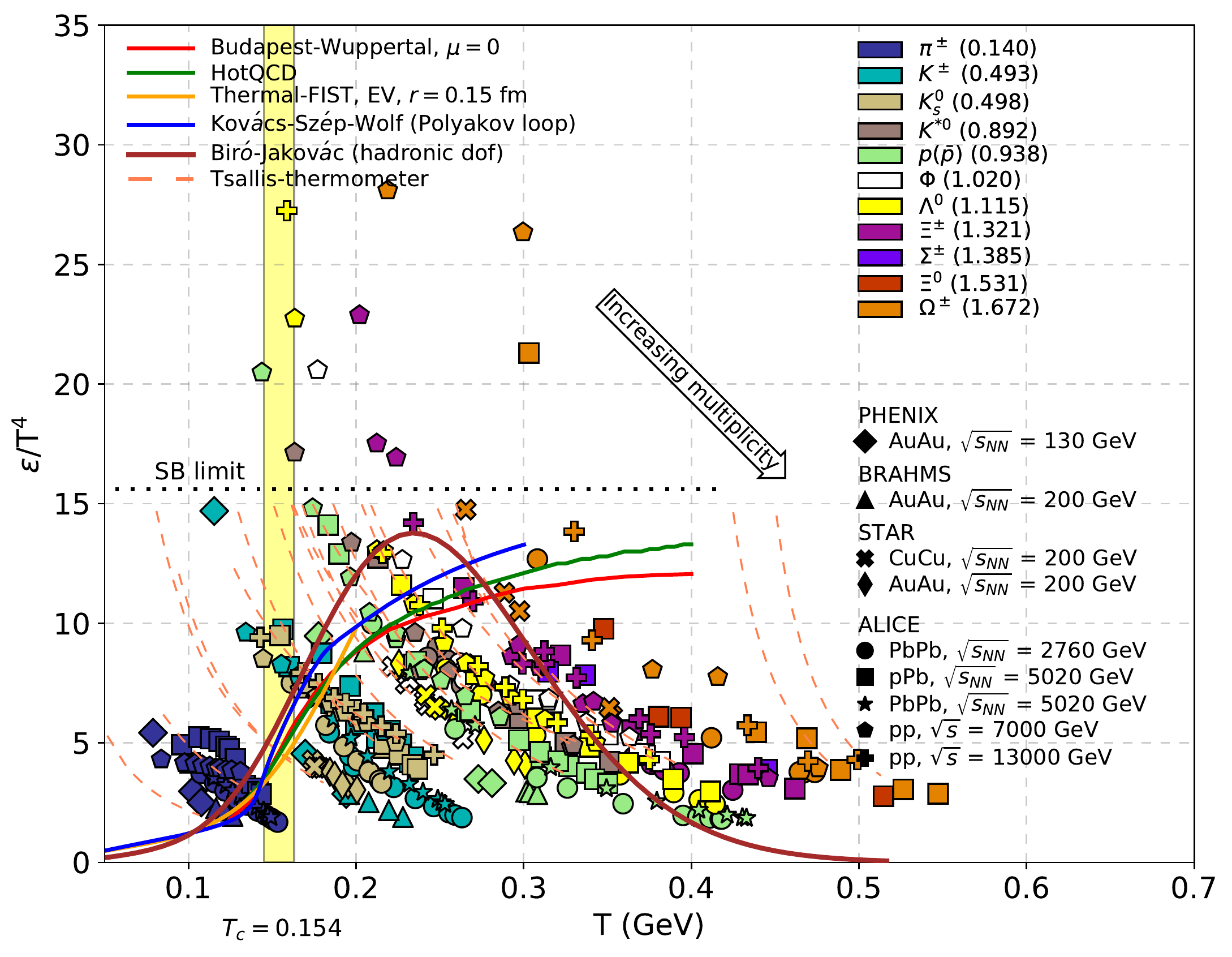}
      \caption{}
      \label{fig:mult_ALL_eT}
    \end{subfigure}
  \caption{The scaled pressure, $P/T^4$ and energy density, $\varepsilon/T^4$ of identified hadrons in pp, pA and AA collisions at various collision energies, in the function of the temperature $T$. The theoretical curves are taken from \cite{Borsanyi:2012cr, Bazavov:2014pvz, Vovchenko:2019pjl, Kovacs:2016juc, Biro:2016one}.}
    \label{fig:mult_ALL_thermo_T1}
  \end{figure}

\begin{figure}[H]
  \centering
      \begin{subfigure}[b]{0.494\textwidth}
        \includegraphics[width=\textwidth]{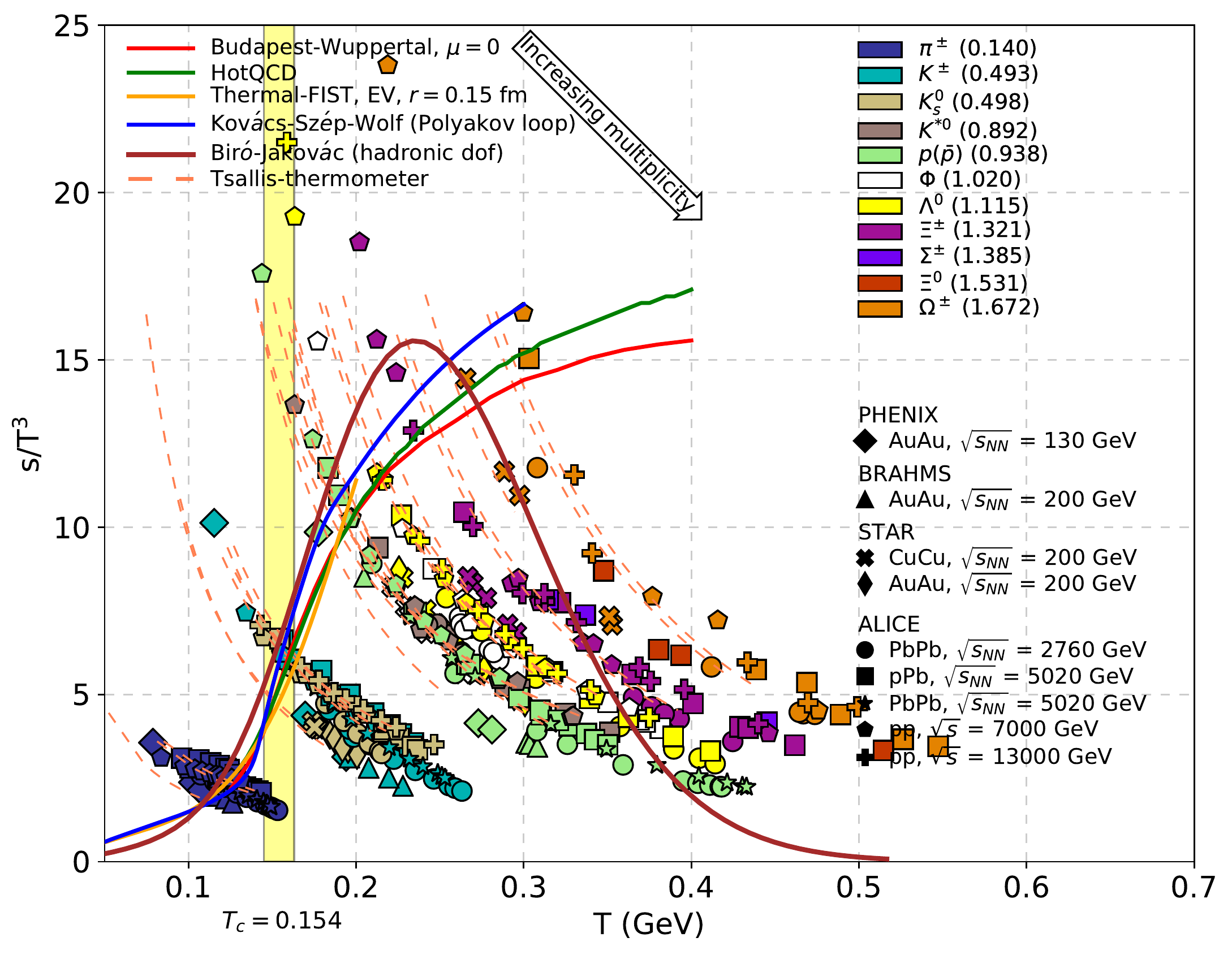}
        \caption{}
        \label{fig:mult_ALL_sT}
      \end{subfigure}
    \begin{subfigure}[b]{0.494\textwidth}
      \includegraphics[width=\textwidth]{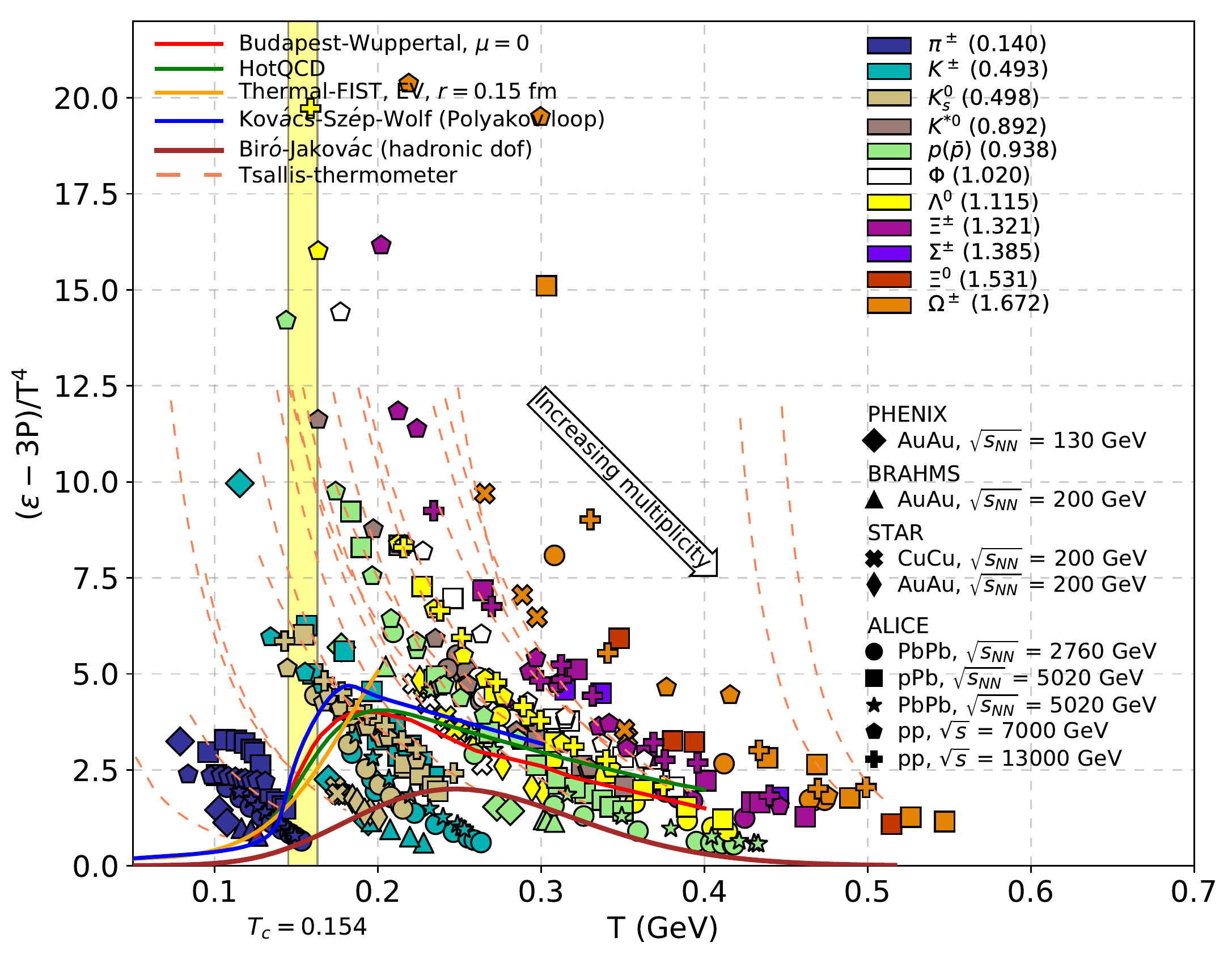}
      \caption{}
      \label{fig:mult_ALL_nT}
      \end{subfigure}
  \caption{The scaled entropy density, $s/T^4$ and interaction measure, $(\varepsilon-3P)/T^4$ of identified hadrons in pp, pA and AA collisions at various collision energies, in the function of the temperature $T$. The theoretical curves are taken from \cite{Borsanyi:2012cr, Bazavov:2014pvz, Vovchenko:2019pjl, Kovacs:2016juc, Biro:2016one}.}
  \label{fig:mult_ALL_thermo_T2}
\end{figure}

In this section the thermodynamical variables are calculated for each hadron species and for each investigated system. The results of the integrations defined by equations \eref{eq:P}-\eref{eq:e} are plotted as functions of the temperature parameter $T$ on \fref{fig:mult_ALL_thermo_T1} and  \fref{fig:mult_ALL_thermo_T2}. For each variable a dimensionless combination scaled with $T$ ($P/T^4$, $\varepsilon/T^4$, $s/T^3$) is plotted. The definition of the colors and markers are the same as on \fref{fig:mult_ALL}.

With solid curves several theoretical models are shown, including lattice QCD calculations (Budapest\,--\,Wuppertal, with red~\cite{Borsanyi:2012cr} and HotQCD, with green~\cite{Bazavov:2014pvz}), a hadron resonance gas model (Thermal-FIST, with orange~\cite{Vovchenko:2019pjl}), a Polyakov constituent quark-meson model (Kov\'acs\,--\,Sz\'ep\,--\,Wolf, with blue~\cite{Kovacs:2016juc}), and an effective field theory with hadronic-like spectral functions in the QGP until temperatures as high as (3-4)$T_c$ (Bir\'o\,--\,Jakov\'ac, with brown~\cite{Biro:2016one}). In this latter approach the thermodynamic quantities can be calculated analytically. Since in our non-extensive approach we started our analysis with the spectra of identified hadrons, in the calculation of the thermodynamic quantities the basic degrees of freedom are the hadrons too. In contrast to that in the models above the QCD matter can be involved only (except in the Bir\'o\,--\,Jakov\'ac model), without separating hadronic degrees of freedom. The definition of 'temperature' is also different in the non-extensive method. Finally, the dashed 'Tsallis-thermometer' curves are our parametrized fits \eref{eq:parametrization_T}-\eref{eq:parametrization_q}. Each curve for each hadron species represent a specific CM energy. Therefore, it is possible to calculate the $\sqrt{s}$~-~$\left<\dd N_{ch}/\dd \eta\right>$ pairs, where our calculations intersect or overlap with the theoretical curves. 

\begin{figure}[H]
  \centering    
  \begin{subfigure}[b]{0.49\textwidth}
    \includegraphics[width=\textwidth]{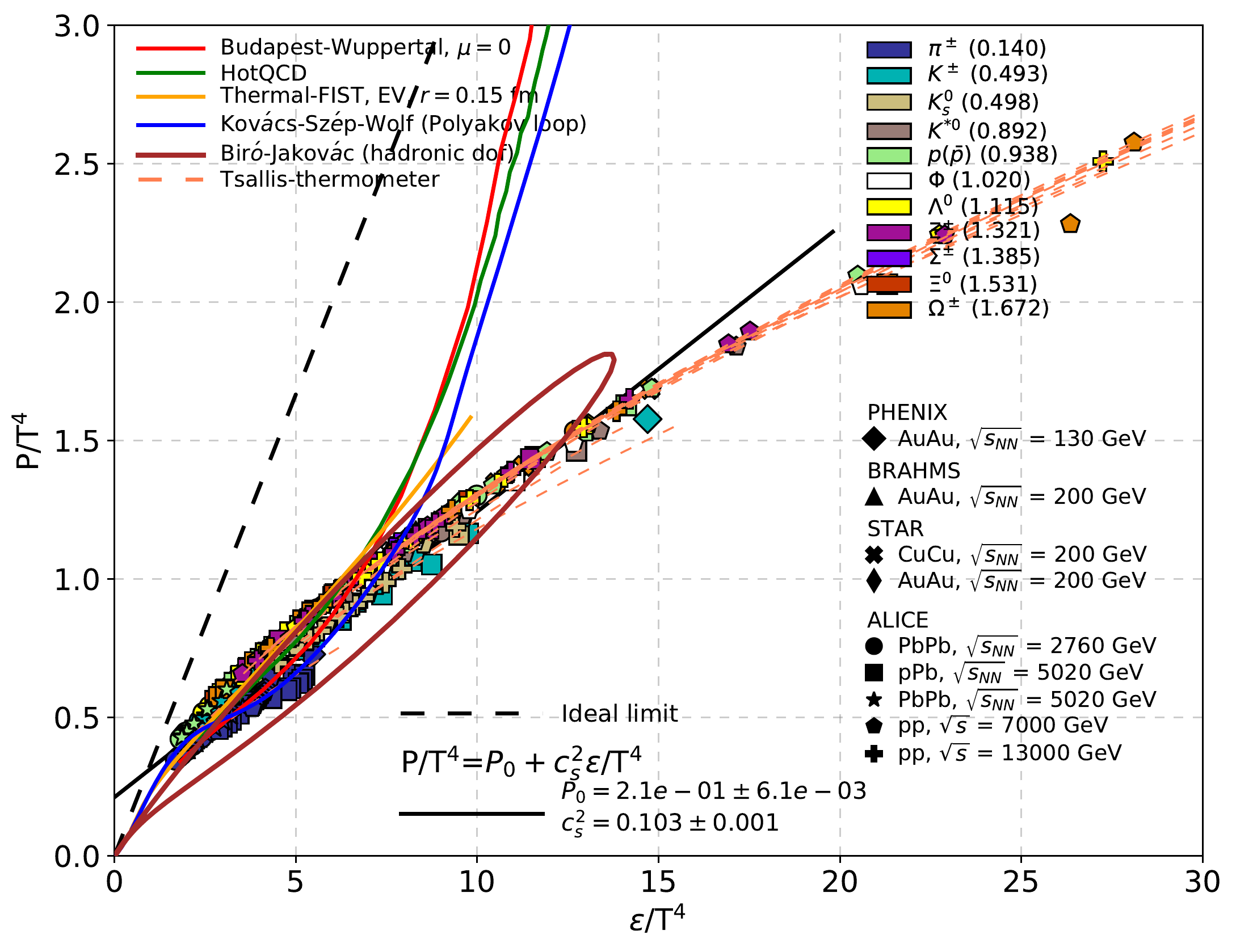}
    \caption{}
    \label{fig:mult_ALL_Pe}
  \end{subfigure}
  \begin{subfigure}[b]{0.49\textwidth}
    \includegraphics[width=\textwidth]{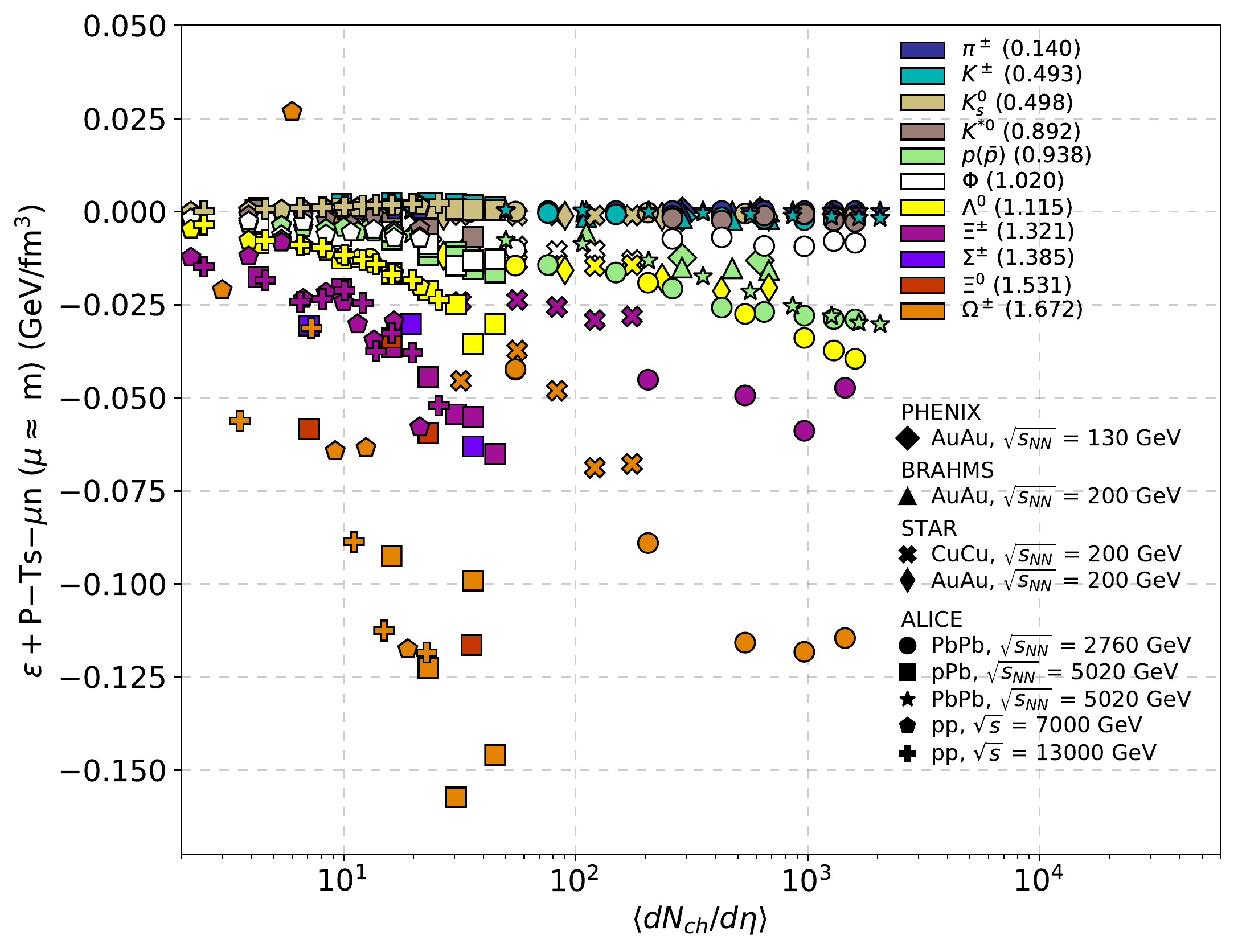}
    \caption{}
    \label{fig:mult_ALL_c}
  \end{subfigure}
  \caption{\textit{Left panel:} The $P(\varepsilon)$ equation of state of identified hadrons in pp, pA and AA collisions at various collision energies. \textit{Right panel:} the examination of thermodynamical consistency (zero is the consistent value).}
  \label{fig:mult_ALL_thermo_3}
\end{figure}

For further validation on the thermodynamical consistency of our approach, in \fref{fig:mult_ALL_Pe} the calculated equation of state, pressure and the energy density values are plotted together, eliminating the differences in the definition of temperature. On \fref{fig:mult_ALL_c} the consistency of the non-extensive statistics can be quantified according to \eref{eq:consistency} explicitly. 

On \fref{fig:mult_ALL_Pe} a quite remarkable observation is that independently of the hadron mass, CM energy and even the multiplicity class, a linear relation holds more or less for each investigated point. A global fit for all points results in a slope parameter, which for ideal gases gives the speed of sound (solid black line):
\begin{equation}
  c_s^2 = \left.\frac{\partial P}{\partial \varepsilon}\right|_{V}\approx 0.103\pm0.001.
\end{equation}
This result, especially with the data points in the low $P/T^4$ and $\varepsilon/T^4$ region (representing the high multiplicity classes) is close to the usual speed of sound calculations and measurements~\cite{Gao:2013dca, Khuntia:2016ikm, Borsanyi:2012cr}. Indeed, by comparing our results with the theoretical models, we can ascertain that most of our calculated points overlap with the theoretical equation of state curves. The ideal radiation limit $c_s^2=1/3$ is indicated by the dashed line.

On \fref{fig:mult_ALL_c} the thermodynamical consistency is tested. For the lighter hadrons like pions and kaons the deviation is very small, $\mathcal{O}(1\%)$, which validates our approach. Data points with larger deviation (5-15\%) stand typically for the heavier multistrange hadrons like $\Omega$, $\Xi$ and $\Sigma$. It might be an indication that the $\mu\approx m$ assumption is no more valid for larger masses and the strangeness content must be taken into account. At the same time the validity of these fits is lesser due to the lack of large high-$p_T$ tails.

\section{Discussion on the role of Tsallis-thermometer}
\label{sec:discussion}

Based on this detailed study the non-extensive approach leads to coherent results. A novel feature is the $q\neq 1$ conclusion from identified hadron spectra. It encodes the dependency on the size of the collisional system through multiplicity fluctuations. In this section we calibrate our 'Tsallis-thermometer' and interpret the grouping phenomenon in the $T$~-~$(q-1)$ parameter space. As already \fref{fig:Tvsqfit2} suggests, the $q$ and $T$ parameters are crowding in a specific region for all hadrons and systems. 

\begin{figure}
  \centering
  
  \begin{subfigure}[b]{0.49\textwidth}
    \includegraphics[width=\textwidth]{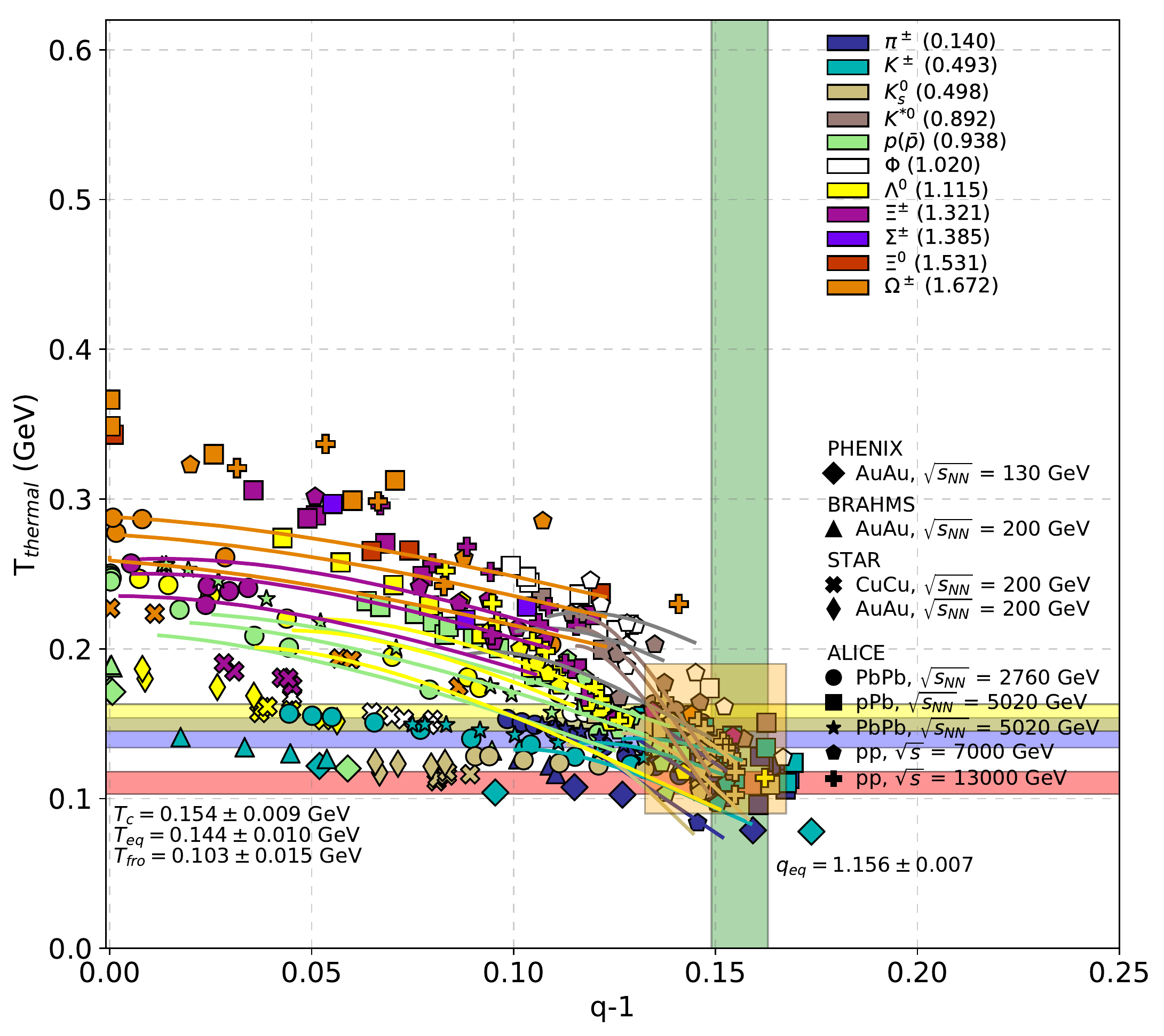}
  \end{subfigure}
  \caption{The temperature parameters corrected for the transverse flow effect apart from the lightest hadrons.}
  \label{fig:flowcorrected}
\end{figure}

In \fref{fig:flowcorrected} the thermal temperature values are shown, resulted from the flow-correction of the spectral temperatures for heavy hadron species:
\begin{equation}
    T_{thermal}=T\sqrt{\frac{1-v_t}{1+v_t}},
\end{equation}
where the transverse flow velocity, $v_t$ is taken from \eref{eq:flowfit}.
The yellow band marks the lattice QCD crossover region, $T_c=0.154\pm0.009$ GeV~\cite{Bazavov:2014pvz}. The red band marks the kinetic freeze-out region, $T_{fro}=0.103\pm0.015$ GeV,  obtained in \sref{sec:flow}, while the green and blue bands indicate the regions where the parameters converge to. The solid lines represent the flow-corrected calculations using the parametrizations \eref{eq:parametrization_q}-\eref{eq:parametrization_T}, in three specific CM energies: $\sqrt{s}\in\{2760, 7000, 13000\}$ GeV.

Our earlier conjecture, that there is a strong grouping of points into a specific area (orange region), seems  confirmed. There is an area where all the different hadron temperatures meet -- and it lies close to the lattice QCD crossover temperature, however, with $q\neq 1$. This region of the non-extensive statistics agrees with theoretical calculations, which indicates the presence of hot QCD matter just before the hadronization. In other words, this means that those colliding systems that are resulting in hadron spectra with Tsallis-parameters close to $T\approx T_{eq}=0.144\pm0.010$ GeV and $q\approx q_{eq}=1.156\pm0.007$ should originate from a previous quark-gluon plasma state. This QGP does certainly not follow an equilibrium Boltzmann\,--\,Gibbs statistics. 

Our parametrizations depending on the center-of-mass energy, multiplicity and the given hadron mass, we can identify those kinematical setups, where one is closest to this data crowding range, at the QGP before hadronization. 
Along with our calculations presented in the previous section, we propose to view the corresponding collision events as best candidates for a previous QGP formation.
Taken into account this specific region of the Tsallis-thermometer, the $\sqrt{s}$ and $\left<\dd N_{ch}/\dd \eta\right>$ values can be obtained from our model. This suggests that for light hadrons at $\sqrt{s}\gtrsim200$ GeV CM energies $\left<\dd N_{ch}/\dd \eta\right>\gtrsim1500$ is the the region, where hadron formation from a QGP-like system with strong correlation is expected. However, it is interesting to see, that this regime can be also reached in small systems like pp collisions at $\sqrt{s}\gtrsim7000$ GeV with $\left<\dd N_{ch}/\dd \eta\right>\gtrsim130$, or even with $\left<\dd N_{ch}/\dd \eta\right>\gtrsim$90 at $\sqrt{s}\gtrsim13000$ GeV.

\section{Summary}
\label{sec:summary}

In this paper we investigated spectra of identified hadrons stemming from pp, pA and AA collisions within the thermodynamically consistent framework of non-extensive statistics. We extracted the Tsallis\,--\,Pareto parameters from a large variety of publicly available experimental data and studied the effect of various collision energies, hadron masses and multiplicity classes on the non-extensive parameters $T$ and $q$. We confirmed correlations between them.

For minimum bias pp collisions we studied the center-of-mass energy dependence of charged pions, kaons and protons in the range of $\sqrt{s}=62.7$-$7000$ GeV. We confirmed our earlier conclusions, namely that both the non-extensivity parameter $q$ and the Tsallis temperature $T$ are increasing with the CM energy in a logarithmic way as $\sim \ln{\sqrt{s}}$, inspired by pQCD. Furthermore, based on the experimentally confirmed observation that multiplicity distributions of the final state charged particles stemming from high-energy collisions can be characterized with negative binomial distributions, we pointed out that such correlations can be related to the multiplicity dependency of the parameters. Since the non-extensive parameters are connected to multiplicity fluctuations, a further linear relation between $T$ and $(q-1)$ can be established. Moreover, depending on the average multiplicity, a slope with negative sign also can be accommodated. The resulted 'multiplicity-lines' are intersecting the '$\sqrt{s}$-lines', suggesting that the latter correspond to multiplicity-averaged minimum bias values.

By studying the spectra of identified hadrons in different multiplicity classes we verified this assumption and showed that the multiplicity has an observable effect on the parameters. Due to the availability of high quality experimental data we could analyze the $p_T$ distribution of eleven identified hadron species, ranging from the lightest charged pions up to the heaviest $\Omega$ baryon, from the very low multiplicity of $\left<\dd N_{ch}/\dd \eta\right>=2.2$ measured in pp collisions, up to the very high multiplicity of  $\left<\dd N_{ch}/\dd \eta\right>=2047$ measured in PbPb collisions, in the $\sqrt{s_{NN}}$ range of 130-13000 GeV. With a na\"ive extension of the pQCD inspired parametrizations of the Tsallis\,--\,Pareto parameters a prediction of identified hadron yields was given at any center-of-mass energy and any centrality.

The previously observed mass hierarchy of the parameters was confirmed for $T$, $q$ and the normalizing factor $A$. In the case of the temperature $T$, assuming a common kinetic freeze-out temperature for each investigated hadron species radial flow was observed. We extracted a kinetic freeze-out temperature, $T_{fro}$ and the average transverse radial flow velocity, $\left<v_t\right>$ for each multiplicity class and center-of-mass energy, using the non-relativistic flow approximation. For a certain part of data, this should be improved in the future. We found that the results obtained via the non-extensive approach are compatible with Boltzmann\,--\,Gibbs-based models.

In our analysis we applied the thermodynamical formulation of non-extensive statistics, so the familiar thermodynamical variables could have been availed. We compared our results, based on hadronic degrees of freedom, with other theoretical models based on quark-gluon and quark-gluon+hadronic degrees of freedom. A global linear fit over all the calculated pressure and energy density values resulted in a slope parameter belonging to the speed of sound, $c_s^2=1/6$, supporting the softest point hadronization model~\cite{Hung:1994eq, Biro:1999eh}.

The most prominent result of the present study is the calibration of the 'Tsallis-thermometer': by investigating the $T$~-~$(q-1)$ parameter space we identified a strong grouping. All hadron data tend to a specific $T_{eq}$ and $q_{eq}$ value. This non-trivial observation was further strengthened by studying the multiplicity dependence: at low multiplicities all parameters are converging to $T_{eq}=0.144\pm0.010$ GeV and $q_{eq}=1.156\pm0.007$. This conclusion was verified by comparing our thermodynamical variables with the lattice QCD and Jakov\'ac\,--\,Bir\'o model curves. They suggest that the region, where the non-extensive temperature is the same as the color deconfinement temperature is eventually not in the $q\rightarrow 1$ Boltzmann\,--\,Gibbs limit, but it is deep in the sector of non-extensivity.

With this interpretation we were able to propose kinematical configurations where the final state hadrons are expected to originate from a previously present strongly interacting QCD matter. Most importantly, our analysis suggests that at LHC energies the event multiplicity can be as low as $\left<\dd N_{ch}/\dd \eta\right>\sim 100$, accessible for small systems.

\ack
The research was supported by the Hungarian National Research, Development and Innovation Office (NKFIH) under the contract numbers OTKA K120660, K123815 and NKFIH 2019-2.1.11-T\'ET-2019-00078, 2019-2.1.11-T\'ET-2019-00050, the Wigner GPU Laboratory and THOR Cost Action CA15213.

\appendix

\section{Investigated datasets}
\label{apx:expdata}
\setcounter{section}{1}

\begin{table}[H]
  {\footnotesize
  \caption{The kinematical regions of the investigated public experimental datasets measured in minimum bias pp collisions at various energies and the references for their publications.}
  \label{tab:exp_sets_1}
    \begin{tabular}{@{}cccccc}
      \br
  System,  & \multirow{2}{*}{$\eta$ or $y$ range} & \multirow{2}{*}{Hadron} & $\left<\frac{dN_{ch}}{d\eta}\right>$ range, num. & $p_T$ range  & \multirow{2}{*}{Reference} \\
  c.m. energy (GeV) & & & of mult. classes & (GeV) & \\
  \mr
  pp, 62.7 & $|\eta|<0.35$ & $\pi^\pm$    & Min. bias & [0.3; 2.9] & \cite{Adare:2011vy} \\
  && $K^\pm$      &           & [0.4; 2.0] & \cite{Adare:2011vy} \\
  && $p(\bar{p})$ &           & [0.6; 4.0] & \cite{Adare:2011vy} \\
  pp, 200 & $|\eta|<0.35$  & $\pi^\pm$    & Min. bias & [0.3; 3.0] & \cite{Adare:2011vy} \\
  && $K^\pm$      &           & [0.4; 2.0] & \cite{Adare:2011vy} \\
  && $p(\bar{p})$ &           & [0.5; 5.0] & \cite{Adare:2011vy} \\
  pp, 900 & $|y|<0.9$
   & $\pi^\pm$    & Min. bias & [0.1; 2.6] & \cite{Aamodt:2011zj} \\
  && $K^\pm$      &           & [0.2; 2.4] & \cite{Aamodt:2011zj} \\
  && $p(\bar{p})$ &           & [0.35; 2.4] & \cite{Aamodt:2011zj} \\
  &$|y|<1.0$      
  & $\pi^\pm$    & Min. bias & [0.1; 1.2] & \cite{Chatrchyan:2012qb} \\
  && $K^\pm$      &           & [0.2; 1.05] & \cite{Chatrchyan:2012qb} \\
  && $p(\bar{p})$ &           & [0.35; 1.7] & \cite{Chatrchyan:2012qb} \\
  pp, 2760 & $|\eta|<0.8$  & $\pi^\pm$    & Min. bias & [0.1; 20.0] & \cite{Abelev:2014laa} \\
  && $K^\pm$      &           & [0.2; 20.0] & \cite{Abelev:2014laa} \\
  && $p(\bar{p})$ &           & [0.3; 20.0] & \cite{Abelev:2014laa} \\
  &$|y|<1.0$      
  & $\pi^\pm$    & Min. bias & [0.1; 1.2] & \cite{Chatrchyan:2012qb} \\
  && $K^\pm$      &           & [0.2; 1.05] & \cite{Chatrchyan:2012qb} \\
  && $p(\bar{p})$ &           & [0.35; 1.7] & \cite{Chatrchyan:2012qb} \\
  pp, 7000 & $|y|<0.5$     & $\pi^\pm$    & Min. bias & [0.1; 3.0]  & \cite{Adam:2015qaa} \\
  && $K^\pm$      &           & [0.2; 6.0]  & \cite{Adam:2015qaa} \\
  && $p(\bar{p})$ &           & [0.3; 6.0] & \cite{Adam:2015qaa} \\
  &$|y|<1.0$      
 & $\pi^\pm$    & Min. bias & [0.1; 1.2] & \cite{Chatrchyan:2012qb} \\
&& $K^\pm$      &           & [0.2; 1.05] & \cite{Chatrchyan:2012qb} \\
&& $p(\bar{p})$ &           & [0.35; 1.7] & \cite{Chatrchyan:2012qb} \\
    \br
  \end{tabular}
}
\end{table}

\begin{table} [H]
  {\footnotesize
\caption{The kinematical regions of the investigated public experimental datasets measured with various centrality/multiplicity selections and at various energies and the references for their publications.}
\label{tab:exp_sets_2}
  \begin{tabular}{@{}cccccc}
    \br
  System,  & \multirow{2}{*}{$\eta$ or $y$ range} & \multirow{2}{*}{Hadron} & $\left<\frac{dN_{ch}}{d\eta}\right>$ range, num. & $p_T$ range  & \multirow{2}{*}{Reference} \\
  c.m. energy (GeV) & & & of mult. classes & (GeV) & \\
  \mr
  AuAu, 130 & $|\eta|<0.35$& $\pi^\pm$    & [21.3; 622], 3 & [0.25; 2.2] & \cite{Adcox:2001mf} \\
  && $K^\pm$      &                & [0.45; 1.65]      & \cite{Adcox:2001mf} \\
  && $p(\bar{p})$ &                & [0.55; 3.42]      & \cite{Adcox:2001mf} \\
  CuCu, 200 & $|y|<0.5$  & $K^0_s$    & [32; 175], 5 & [0.5; 9.0] & \cite{Agakishiev:2011ar} \\
   &  & $\Lambda^0$    &  & [0.5; 7.0] & \cite{Agakishiev:2011ar} \\
   &  & $\Xi^\pm$    &  & [0.7; 6.0] & \cite{Agakishiev:2011ar} \\
   &  & $\Omega^\pm$    &  & [1.0; 4.5] & \cite{Agakishiev:2011ar} \\
   &  & $\Phi$    & [24, 175], 6 & [0.45; 4.5] & \cite{Abelev:2008zk} \\
  AuAu, 200 & $|y|<0.2$    & $\pi^\pm$    & [111; 680], 3 & [0.2; 2.0] & \cite{Arsene:2005mr} \\
  && $K^\pm$      &                & [0.4; 2.0]      & \cite{Arsene:2005mr} \\
  && $p(\bar{p})$ &                & [0.3; 3.0]      & \cite{Arsene:2005mr} \\
  & $|y|<0.5$ & $K^0_s$      &  [27, 680], 5        & [0.5; 9.0]      & \cite{Agakishiev:2011ar} \\
  &           & $\Lambda^0$      &                  & [0.5; 8.0]      & \cite{Agakishiev:2011ar} \\
  PbPb, 2760 & $|y|<0.5$   & $\pi^\pm$    & [13.4; 1601], 10 & [0.1; 3.0] & \cite{Abelev:2014laa, Abelev:2013vea} \\
  && $K^\pm$      &           & [0.2; 3.0]        & \cite{Abelev:2014laa, Abelev:2013vea} \\
  && $K^0_s$      & [55; 1601], 7 & [0.4; 12.0]   & \cite{Abelev:2013xaa} \\
  && $K^{*0}$     & [261; 1601], 6 & [0.3; 20.0]   & \cite{Adam:2017zbf} \\
  && $p(\bar{p})$ &           & [0.3; 4.6]        & \cite{Abelev:2014laa, Abelev:2013vea} \\
  && $\Lambda^0$ &              & [0.6; 12.0]  & \cite{Abelev:2013xaa} \\
  && $\Phi$       &             & [0.5; 21.0]  & \cite{Adam:2017zbf} \\
  && $\Xi^\pm$      & [55; 1601], 5  & [0.6; 8.0]   & \cite{ABELEV:2013zaa} \\
  && $\Omega^\pm$   &               & [1.2; 7.0]  & \cite{ABELEV:2013zaa} \\
  pPb, 5020 & $-0.5<|y|<0.0$& $\pi^\pm$    & [4.3; 45], 7 & [0.1; 20.0]   & \cite{Adam:2016dau} \\
  && $K^\pm$      &              & [0.2; 20.0]   & \cite{Adam:2016dau} \\
  && $K^{*0}$     & [4.3; 45], 5 & [0.0; 16.0]   & \cite{Adam:2016bpr} \\
  && $p(\bar{p})$ &              & [0.35; 20.0]  & \cite{Adam:2016dau} \\
  && $\Phi$       &              & [0.4; 20.0]  & \cite{Adam:2016bpr} \\
  && $\Xi^0$        & [7.1; 35.6], 4  & [0.8; 8.0]   & \cite{Adamova:2017elh} \\
  && $\Sigma^\pm$   & [7.1; 35.6], 3 & [1.0; 6.0]  & \cite{Adamova:2017elh} \\
  && $\Xi^\pm$      & [4.3; 45], 7  & [0.6; 7.2]   & \cite{Adam:2015vsf} \\
  && $\Omega^\pm$   &               & [0.8; 5.0]  & \cite{Adam:2015vsf} \\
  & $0.0<|y|<0.5$ & $\pi^\pm$    & [4.3; 45], 7 & [0.1; 3.0]   & \cite{Abelev:2013haa} \\
  && $K^\pm$      &              & [0.2; 2.4]   & \cite{Abelev:2013haa} \\
  && $K^0_s$      &              & [0.0; 8.0]   & \cite{Abelev:2013haa} \\
  && $p(\bar{p})$ &              & [0.3; 4.0]  & \cite{Abelev:2013haa} \\
  && $\Lambda^0$ &              & [0.6; 8.0]  & \cite{Abelev:2013haa} \\
  \br
  \end{tabular}
}
\end{table}

\setcounter{table}{1}

\begin{table} [H]
  {\footnotesize
\caption{Continued from previous page.}
  \begin{tabular}{@{}cccccc}
    \br
  System,  & \multirow{2}{*}{$\eta$ or $y$ range} & \multirow{2}{*}{Hadron} & $\left<\frac{dN_{ch}}{d\eta}\right>$ range, num. & $p_T$ range  & \multirow{2}{*}{Reference} \\
  c.m. energy (GeV) & & & of mult. classes & (GeV) & \\
  \mr
  PbPb, 5020 &  $|y|<0.5$  & $\pi^\pm$    & [19.5; 2047], 10 & [0.1; 10.0] & \cite{Acharya:2019yoi} \\
  && $K^\pm$      &           & [0.1; 10.0]        & \cite{Acharya:2019yoi} \\
  && $p(\bar{p})$ &           & [0.1; 10.0]        & \cite{Acharya:2019yoi} \\
  pp, 7000 & $|y|<0.5$ & $\pi^\pm$    & [2.2; 21.3], 10 & [0.1; 20.0]   & \cite{Acharya:2018orn} \\
  && $K^\pm$      &              & [0.2; 20.0]   & \cite{Acharya:2018orn} \\
  && $K^0_s$      &   [2.2; 21.3], 10 & [0.0; 12.0]   & \cite{ALICE:2017jyt} \\
  && $K^{*0}$     & [2.2; 21.3], 9 & [0.0; 10.0]   & \cite{Acharya:2018orn} \\
  && $p(\bar{p})$ & [2.2; 21.3], 10 & [0.3; 20.0]  & \cite{Acharya:2018orn} \\
  && $\Phi$       & [2.2; 21.3], 9 & [0.4; 10.0]  & \cite{Acharya:2018orn} \\
  && $\Lambda^0$  & [2.2; 21.3], 10 & [0.4; 8.0]   & \cite{ALICE:2017jyt} \\
  && $\Xi^\pm$    &                   & [0.6; 6.5]   & \cite{ALICE:2017jyt} \\
  && $\Omega^\pm$ & [2.2; 21.3], 5    & [0.9; 5.5]   & \cite{ALICE:2017jyt} \\
  pp, 13000 & $|y|<0.5$ &  $K^0_s$      &   [2.52; 25.72], 10 & [0.0; 12.0]   & \cite{Acharya:2019kyh} \\
  && $\Lambda^0$  &                   & [0.4; 8.0]   & \cite{Acharya:2019kyh} \\
  && $\Xi^\pm$    &                   & [0.6; 6.5]   & \cite{Acharya:2019kyh} \\
  && $\Omega^\pm$ & [3.58; 22.8], 5    & [0.9; 5.5]   & \cite{Acharya:2019kyh} \\
  \br
  \end{tabular}
}
\end{table}

\section{Fitted parameters: spectra}
\label{apx:parameters}

  \input params/pp_62.7.tex
  \input params/pp_200.tex
  \input params/pp_900.tex
  \input params/pp_2760.tex
  \input params/pp_7000.tex
  \input params/AuAu_013_pipm.tex
  \input params/AuAu_013_Kpm.tex
  \input params/AuAu_013_ppm.tex
  \input params/CuCu_02_K0s.tex
  \input params/CuCu_02_lambdapm.tex
  \input params/CuCu_02_omegapm.tex
  \input params/CuCu_02_Phi.tex
  \input params/CuCu_02_xipm.tex
  \input params/AuAu_02_pipm.tex
  \input params/AuAu_02_Kpm.tex
  \input params/AuAu_02_ppm.tex
  \input params/PbPb_2_pipm.tex
  \input params/PbPb_2_Kpm.tex
  \input params/PbPb_2_K0s.tex
  \input params/PbPb_2_Kstar.tex
  \input params/PbPb_2_Phi.tex
  \input params/PbPb_2_ppm.tex
  \input params/PbPb_2_Lambda.tex
  \input params/PbPb_2_xipm.tex
  \input params/PbPb_2_omegapm.tex
  \input params/pPb_5_pipm.tex
  \input params/pPb_5_pipm_2.tex
  \input params/pPb_5_Kpm.tex
  \input params/pPb_5_Kpm_2.tex
  \input params/pPb_5_K0s_2.tex
  \input params/pPb_5_Kstar.tex
  \input params/pPb_5_Phi.tex
  \input params/pPb_5_ppm.tex
  \input params/pPb_5_ppm_2.tex
  \input params/pPb_5_lambda_2.tex
  \input params/pPb_5_xipm.tex
  \input params/pPb_5_sigma.tex
  \input params/pPb_5_xi0.tex
  \input params/pPb_5_omega.tex
  \input params/PbPb_5_pipm.tex
  \input params/PbPb_5_Kpm.tex
  \input params/PbPb_5_ppm.tex
  \input params/pp_7m_pipm.tex
  \input params/pp_7m_Kpm.tex
  \input params/pp_7_K0s.tex
  \input params/pp_7m_Kstar.tex
  \input params/pp_7m_ppm.tex
  \input params/pp_7m_phi.tex
  \input params/pp_7_lambda.tex
  \input params/pp_7_xipm.tex
  \input params/pp_7_omega.tex
  \input params/pp_13_K0s.tex
  \input params/pp_13_Lambda.tex
  \input params/pp_13_Xipm.tex
  \input params/pp_13_Omega.tex
  
\section{Fitted parameters: correlations in the $T$~-~$(q-1)$ parameter space}
\label{apx:Tvsqfits}

\begin{table}[H]
  \caption{Fit parameters of \eref{eq:Tvsqfit} on \fref{fig:Tvsqfit2}.}
  \label{tab:Tvsqfit2}
  \begin{indented}
    \item[]
  {\scriptsize
    \begin{tabular}{@{}cccccc}
      \br
  $\sqrt{s_{NN}}$ (GeV) & Hadron & $E$ (GeV) & $\delta^2$ & $E\delta^2$ (GeV) & $\chi^2/ndf$\\
  \mr
  \input params/Tqfits_good.tex
    \br 
  \end{tabular}
}
\end{indented}
\end{table}

\section{Fitted parameters: average event multiplicity scaling}
\label{apx:mult_parameters}

\input params/mult_ALL_T.tex
\input params/mult_ALL_q.tex
\input params/mult_ALL_A.tex

\section*{References}
\bibliographystyle{unsrtnat}

\end{document}

%% file: params/Tqfits_good.tex
AuAu, 130 & $\pi^\pm$ & $0.684\pm0.120$ & $0.275\pm0.031$ & $0.188\pm0.124$ & $0.049/1$\\ 
  & $K^\pm$ & $0.705\pm0.278$ & $0.330\pm0.099$ & $0.232\pm0.295$ & $0.034/1$\\ 
  & $p(\bar{p})$ & $1.639\pm2.003$ & $0.172\pm0.209$ & $0.281\pm2.014$ & $1.160/1$\\ 
 AuAu, 200 & $\pi^\pm$ & $0.800\pm0.195$ & $0.258\pm0.039$ & $0.207\pm0.199$ & $0.323/2$\\ 
  & $K^\pm$ & $0.777\pm0.107$ & $0.312\pm0.039$ & $0.242\pm0.114$ & $0.076/2$\\ 
  & $p(\bar{p})$ & $1.920\pm1.283$ & $0.159\pm0.106$ & $0.304\pm1.288$ & $3.153/2$\\ 
  & $K^0_s$ & $0.887\pm0.175$ & $0.290\pm0.042$ & $0.257\pm0.180$ & $1.109/3$\\ 
CuCu, 200 & $K^0_s$ & $0.605\pm0.287$ & $0.375\pm0.138$ & $0.227\pm0.318$ & $1.334/3$\\ 
  & $\Phi$ & $1.168\pm0.195$ & $0.272\pm0.034$ & $0.318\pm0.198$ & $1.284/4$\\ 
  & $\Lambda^0$ & $1.379\pm0.065$ & $0.219\pm0.008$ & $0.302\pm0.066$ & $0.116/3$\\ 
  & $\Xi^{\pm}$ & $1.510\pm0.338$ & $0.222\pm0.041$ & $0.335\pm0.340$ & $0.975/3$\\ 
  & $\Omega^\pm$ & $1.392\pm0.072$ & $0.265\pm0.012$ & $0.369\pm0.073$ & $0.151/3$\\ 
  & $\Lambda^0$ & $1.508\pm0.088$ & $0.207\pm0.010$ & $0.312\pm0.088$ & $0.616/3$\\ 
 PbPb, 2760 & $\pi^\pm$ & $0.969\pm0.037$ & $0.260\pm0.006$ & $0.252\pm0.038$ & $6.644/8$\\ 
  & $K^\pm$ & $0.956\pm0.016$ & $0.322\pm0.004$ & $0.308\pm0.016$ & $2.391/8$\\ 
  & $K^0_s$ & $0.901\pm0.060$ & $0.333\pm0.014$ & $0.300\pm0.062$ & $0.866/5$\\ 
  & $K^{*0}$ & $1.349\pm0.134$ & $0.326\pm0.021$ & $0.440\pm0.135$ & $0.219/4$\\ 
  & $p(\bar{p})$ & $1.891\pm0.238$ & $0.213\pm0.026$ & $0.402\pm0.240$ & $469.739/8$\\ 
  & $\Phi$ & $1.100\pm0.159$ & $0.364\pm0.037$ & $0.401\pm0.163$ & $0.312/5$\\ 
  & $\Lambda^0$ & $1.971\pm0.091$ & $0.222\pm0.007$ & $0.438\pm0.091$ & $2.150/5$\\ 
  & $\Xi^{\pm}$ & $1.902\pm0.079$ & $0.227\pm0.009$ & $0.431\pm0.080$ & $1.122/3$\\ 
  & $\Omega^\pm$ & $2.123\pm0.354$ & $0.226\pm0.035$ & $0.480\pm0.356$ & $1.671/3$\\ 
 pPb, 5020 & $\pi^\pm$ & $3.300\pm0.394$ & $0.183\pm0.005$ & $0.603\pm0.394$ & $83.023/5$\\ 
  & $\pi^\pm$ (p) & $3.495\pm2.161$ & $0.189\pm0.019$ & $0.659\pm2.161$ & $208.674/5$\\ 
  & $K^\pm$ & $2.773\pm0.268$ & $0.214\pm0.007$ & $0.594\pm0.268$ & $81.405/5$\\ 
  & $K^\pm$ (p) & $1.748\pm0.312$ & $0.270\pm0.021$ & $0.471\pm0.313$ & $78.193/5$\\ 
  & $K^0_s$ (p) & $2.080\pm0.190$ & $0.237\pm0.009$ & $0.494\pm0.190$ & $42.093/5$\\ 
  & $K^{*0}$ & $2.509\pm0.137$ & $0.245\pm0.006$ & $0.615\pm0.137$ & $2.638/3$\\ 
  & $p(\bar{p})$ & $2.645\pm0.076$ & $0.205\pm0.003$ & $0.542\pm0.076$ & $9.636/5$\\ 
  & $p(\bar{p})$ (p) & $2.185\pm0.122$ & $0.229\pm0.007$ & $0.499\pm0.122$ & $33.095/5$\\ 
  & $\Phi$ & $2.658\pm0.133$ & $0.243\pm0.006$ & $0.647\pm0.133$ & $3.867/5$\\ 
  & $\Lambda^0$ (p) & $2.565\pm0.055$ & $0.210\pm0.002$ & $0.540\pm0.055$ & $1.615/5$\\ 
  & $\Xi^{\pm}$ & $2.143\pm0.130$ & $0.249\pm0.010$ & $0.533\pm0.131$ & $10.502/5$\\ 
  & $\Sigma^{\pm}$ & $2.363\pm0.060$ & $0.238\pm0.005$ & $0.562\pm0.060$ & $0.008/1$\\ 
  & $\Xi^0$ & $2.104\pm0.608$ & $0.260\pm0.055$ & $0.546\pm0.610$ & $5.689/2$\\ 
  & $\Omega^\pm$ & $1.449\pm0.227$ & $0.383\pm0.058$ & $0.555\pm0.234$ & $34.285/5$\\ 
 PbPb, 5020 & $\pi^\pm$ & $0.823\pm0.097$ & $0.288\pm0.017$ & $0.237\pm0.098$ & $4.547/8$\\ 
  & $K^\pm$ & $1.125\pm0.040$ & $0.305\pm0.006$ & $0.343\pm0.041$ & $7.049/8$\\ 
  & $p(\bar{p})$ & $1.949\pm0.052$ & $0.234\pm0.004$ & $0.456\pm0.052$ & $7.090/8$\\ 
 pp, 7000 & $\pi^\pm$ & $0.130\pm3.513$ & $1.000\pm24.746$ & $0.130\pm24.994$ & $1398.474/8$\\ 
  & $K^\pm$ & $5.407\pm3.352$ & $0.183\pm0.019$ & $0.989\pm3.352$ & $3285.512/8$\\ 
  & $K^0_s$ & $6.255\pm1.415$ & $0.177\pm0.007$ & $1.104\pm1.415$ & $506.698/8$\\ 
  & $K^{*0}$ & $4.153\pm0.297$ & $0.194\pm0.004$ & $0.805\pm0.297$ & $10.288/7$\\ 
  & $p(\bar{p})$ & $5.259\pm0.403$ & $0.167\pm0.003$ & $0.877\pm0.403$ & $116.540/8$\\ 
  & $\Phi$ & $3.599\pm0.239$ & $0.214\pm0.005$ & $0.770\pm0.239$ & $15.319/7$\\ 
  & $\Lambda^0$ & $3.246\pm0.058$ & $0.190\pm0.001$ & $0.618\pm0.058$ & $3.299/8$\\ 
  & $\Xi^{\pm}$ & $3.126\pm0.137$ & $0.195\pm0.004$ & $0.609\pm0.138$ & $8.680/8$\\ 
  & $\Omega^\pm$ & $1.664\pm0.320$ & $0.303\pm0.045$ & $0.504\pm0.323$ & $2.699/3$\\ 
 pp, 13000 & $K^0_s$ & $6.177\pm0.962$ & $0.182\pm0.005$ & $1.126\pm0.962$ & $470.847/8$\\ 
  & $\Lambda^0$ & $3.604\pm0.117$ & $0.187\pm0.002$ & $0.675\pm0.117$ & $11.581/8$\\ 
  & $\Xi^{\pm}$ & $3.024\pm0.144$ & $0.209\pm0.005$ & $0.633\pm0.145$ & $8.800/8$\\ 
  & $\Omega^\pm$ & $1.592\pm0.480$ & $0.364\pm0.089$ & $0.579\pm0.488$ & $2.200/3$\\ 
 